\documentclass[fleqn,usenatbib]{mnras}

\newcommand{\orcid}[1]{\href{http://orcid.org/#1}{\includegraphics[width=9pt]{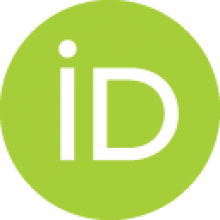}}}
\usepackage{newtxtext,newtxmath}

\usepackage[T1]{fontenc}
\DeclareRobustCommand{\VAN}[3]{#2}
\let\VANthebibliography\thebibliography
\def\thebibliography{\DeclareRobustCommand{\VAN}[3]{##3}\VANthebibliography}

\usepackage{graphicx}	
\usepackage{amsmath}	
\usepackage{amssymb}	
\usepackage{algorithm2e}
\usepackage{longtable,verbatim}
\usepackage{soul}       
\usepackage[theme=default-plain, usecourier=false]{jlcode} 

\newcommand{\pdjl}{\texttt{Photodynamics.jl}\xspace}
\newcommand{\nbg}{\texttt{NbodyGradient.jl}\xspace}
\newcommand{\limbdark}{\texttt{Limbdark.jl}\xspace}
\newcommand{\photodynam}{\texttt{photodynam}\xspace}

\newcommand{\nbeta}{\ensuremath{\boldsymbol{\eta}}}
\newcommand{\nbq}{\ensuremath{\mathbf{q}}}
\newcommand{\der}[2]{\ensuremath{\frac{\partial{#1}}{\partial{#2}}}}
\DeclareMathAlphabet{\mathbfcal}{OMS}{cmsy}{b}{n}

\title[Photodynamics]{A differentiable N-body code for transit timing and dynamical modelling -
II. Photodynamics}

\author[Z. Langford et al.]{
Zachary Langford \orcid{0000-0001-7574-4440}$^{1,2}$\thanks{E-mail: langfzac@sas.upenn.edu} and Eric Agol \orcid{0000-0002-0802-9145}$^{2}$ 
\\
$^{1}${Department of Physics and Astronomy, University of Pennsylvania, Philadelphia, PA, 19104, USA}\\
$^{2}${Astronomy Department and Virtual Planetary Laboratory, University of Washington, Seattle, WA 98195, USA}\\
}

\date{Accepted 2025 April 4. Received 2025 March 28; in original form 2023 November 15}

\pubyear{2025}

\begin{document}
\label{firstpage}
\pagerange{\pageref{firstpage}--\pageref{lastpage}}

\maketitle

\begin{abstract}
    Exoplanet transits contain substantial information about the architecture of a system. By fitting transit light curves we can extract dynamical parameters and place constraints on the properties of the planets and their host star. Having a well-defined probabilistic model plays a crucial role in making robust measurements of these parameters, and the ability to differentiate the model provides access to more robust inference tools. Gradient-based inference methods can allow for more rapid and accurate sampling of high-dimensional parameter spaces. We present a fully differentiable photodynamical model for multiplanet transit light curves that display transit-timing variations. We model time-integrated exposures, compute the dynamics of a system over the full length of observations, and provide analytic expressions for derivatives of the flux with respect to the dynamical and photometric model parameters. The model has been implemented in the Julia language and is available open-source on GitHub. We demonstrate with a simulated data set that Bayesian inference with the NUTS HMC algorithm, which uses the model gradient, can outperform the affine-invariant (e.g. \texttt{emcee}) MCMC algorithm in CPU time per effective sample, and we find that the relative sampling efficiency improves with the number of model parameters.
\end{abstract}


\section{Introduction}
Discovery and characterization of exoplanet systems often comes down to staring at stars. Watching a star's brightness fluctuate while planets transit the stellar disk enables astronomers to infer orbital architectures, planetary compositions, and stellar parameters \citep[e.g.][]{Seager2003,Ragozzine2010,Fabrycky2014,Winn2015,AgolFabrycky2018}. This method was employed by the CoROT, Kepler, and K2 missions \citep{Auvergne2009,Borucki2010,Howell2014}; is currently finding planets with TESS \citep{Ricker2014}; and will be used by the PLATO mission in the future \citep{Rauer2014}. The transit method has enabled the discovery of thousands of exoplanets to date \citep[e.g.][]{Christiansen2022}, and continues to be a prolific discovery method.

Often times it is sufficient to model the dynamics of these systems as a sequence of Keplerian orbits, in which the transits and eclipses are computed with time-integration of the relative positions of the bodies using Kepler's equation \citep[e.g.][]{Kipping2010,Kipping2011,Southworth_2012,Gazak2012,Eastman_2013,Parviainen2015,Kreidberg2015,Luger2017,Barragan2019,Espinoza2019,exoplanet-code,Gunther2021}. However, with more than two bodies, when dynamical interactions are strong, the relative positions must be computed via \textit{N}-body integration. The interactions cause variations of the orbital elements, which in turn cause the intervals between transits to vary. These transit-timing variations (TTVs) typically become significant when pairs of planets are near mean-motion resonances \citep{Agol2005,Holman2005}. Other transit parameters -- the duration, impact parameter, and depth of transit -- can vary due to dynamical effects as well.

A common approach to modeling TTVs is to measure the times and uncertainties of individual transits by fitting a transit light curve model \citep[e.g.][]{Mandel2002, Gimenez2006, Maxted2016, Short2018}, and to then compute the transit times with an \textit{N}-body model which is fit to the measured times. This divides the modeling up into two steps, and allows one to account for different numbers of bodies in the dynamical modeling (including non-transiting planets).  The transit durations and impact parameters can be measured and fit with the \textit{N}-body model as well if their time variation due to dynamical interactions is significant. 

A disadvantage with this two-step approach is that the posterior distribution of the transit modeling needs to be summarized or approximated and then passed to the dynamical analysis.  This is particularly problematic when the signal-to-noise of individual transits is low, and sometimes biases may arise in this process which cause the TTVs, and hence the planet masses, to be underestimated \citep[e.g.][]{Leleu2023,Judkovsky2023}.  So, an alternative approach is to combine the \textit{N}-body model and transit models into a single "photodynamical" model \citep{Doyle2011,Carter2012}.  This has the advantage that the dynamical model can be fit to all of the data simultaneously, accounting for all dynamical interactions in the photometric light curve rather than fitting the joint posteriors of the transit parameters of each and every transit.  If the dynamical model is sufficiently complete, this can improve the accuracy of the measurements of radius-ratios, mass-ratios, and orbital elements \citep{Leleu2023}.

Photodynamical models are thus designed to directly fit photometric time-series measurements of dynamically active astrophysical systems. In general, this approach can be used to compute light curves of systems where one or more luminous bodies are occulted by others, such as eclipsing binaries, transiting exoplanets, and transiting exomoons.  An additional benefit is that they can model transits which occur simultaneously, and they can account for accelerations during transit, which is particularly important for tertiary eclipses of eclipsing binaries. Thus, numerous photodynamical codes have been developed over the last decade.  An early photodynamical model is the Eclipsing Light Curve (ELC) model developed by \citet{Orosz2000}, which has been widely used to model circumbinary planets \citep[e.g.][]{Orosz2012} as well as KOI-126 \citep{Yenawine2022}. The photodynamical concept was described at the onset of the Kepler era by \citet{Ragozzine2010}, which was followed by the development of \texttt{photodynam}\footnote{\url{https://github.com/dfm/photodynam}\label{fn:photodynam}} written by Josh Carter and Andr\'as P\'al which was used in modeling the triple star system KOI-126 \citep{Carter2011}, the first transiting circumbinary planets, Kepler-16b and Kepler-38b \citep{Doyle2011,Pal2012,Orosz2012}, and the Kepler-36 and Kepler-56 planet systems \citep{Carter2012,Huber2013}. At about the same time, the photodynamical code \texttt{LIGHTCURVEFACTORY} was being developed to model triple stars in Kepler data \citep{Borkovits2012}.  This model has been widely applied to triple star systems in both Kepler and TESS data sets \citep[e.g.][]{Borkovits2018,Borkovits2020,Gaulme2022}. More recently, the PHysics Of Eclipsing BinariEs (PHOEBE) project implemented multiple \textit{N}-body dynamical models to compute light curves of various multistar systems \citep{Prsa2018}.

Since that time, several proprietary and open-source photodynamical models have been developed and applied to multibody systems. The following list is probably not exhaustive as there has been extensive work on these codes in the era of Kepler, K2 and TESS, and there may be earlier codes we are unaware of before this approach became commonplace. Mills, Fabrycky and Ragozzine developed \texttt{PhoDyMM}\footnote{\url{https://github.com/dragozzine/PhoDyMM}} \citep{Mills2016}; 
\citet{Yoffe2021} developed \texttt{PyDynamicaLC}\footnote{\url{https://github.com/avivofir/PyDynamicaLC}} , which is based on \texttt{TTVFast}\footnote{\url{https://github.com/kdeck/TTVFast}} \citep{Deck2014} or \texttt{TTVFaster}\footnote{\url{https://github.com/ericagol/TTVFaster}} \citep{Agol2016}; 
Almenara developed a proprietary code, which has been applied to  K2-19 \citep{Barros2015}, Kepler-138 \citep{Almenara2018}, and Kepler-117 \citep{Almenara_2015};
\citet{Migaszewski2012} developed a proprietary code and applied it to Kepler-11;
\citet{Freudenthal2018} applied their proprietary code to Kepler-9; \texttt{AnalyticLC}\footnote{\url{https://github.com/yair111/AnalyticLC}} from \cite{AnalyticLC,Judkovsky2024} uses an analytic dynamical model for transit light curves with TTVs; \cite{exoplanet-code} developed the \texttt{exoplanet}\footnote{\url{https://github.com/exoplanet-dev/exoplanet}} code which employs automatic-differentiation (auto-diff or AD) for gradients; \texttt{jnkepler}\footnote{\url{https://github.com/kemasuda/jnkepler}} also uses auto-diff for gradients, assuming linear trajectories during transit \citep{Masuda2024}; and \cite{korth2020} developed \texttt{PyTTV} that uses a novel dynamical model, which was the inspiration for this work (see also \citealt{korth2023}). Table \ref{tab:codes} provides a breakdown of the relevant features for a selection of the accessible, open-source codes.

\begin{table*}
\caption{A selection of existing accessible, open-source, photodynamical modeling codes used for transiting exoplanets. Transit models: Mandel-Agol \citep{Mandel2002}, Pál \citep{Pal2012}, and ALFM20 \citep{Agol2020}. Dynamical models: \textit{N}-body in this case simply means some choice of numerical integration for solving the \textit{N}-body problem directly to compute the sky-plane coordinates at the exposure times; \textit{N}-body + Linear is an \textit{N}-body integration to compute the sky-plane position and velocity at the transit mid-point, and then assume a linear trajectory during transit; \textit{N}-body + PK20 is the method outlined in this paper.}
\begin{tabular}{lllcc}
\hline 
Code & Transit model & Dynamics & Gradients & Integrated exposures \\ \hline
\texttt{AnalyticLC}     & Mandel-Agol & Analytic          & $\times$          & $\times$           \\
\texttt{photodynam} & Pál & \textit{N}-body    & $\times$ & $\times$ \\
\texttt{PhoDyMM}          & Pál         & \textit{N}-body + Linear           & $\times$          & \checkmark           \\
\texttt{PyDynamicaLC}     & Various & Analytic \& \textit{N}-body & $\times$          & \checkmark \\
\texttt{exoplanet}        & ALFM20    & Keplerian*        & \checkmark & \checkmark  \\
\texttt{jnkepler}         & ALFM20    & \textit{N}-body + Linear   & \checkmark & \checkmark \\
\texttt{Photodynamics.jl} & ALFM20    & \textit{N}-body + PK20           & \checkmark & \checkmark \\
\hline
\end{tabular} 

\textit{Note}. *\textit{N}-body is mentioned in the docs, but is not currently listed in the public API
\label{tab:codes}
\end{table*}

What nearly all of the preceding photodynamical models lack is a means to quickly and accurately compute the gradients of a light curve model with respect to the initial conditions (exceptions being \texttt{exoplanet} and \texttt{jnkepler}). For the purposes of data-modeling, access to the gradient of the model with respect to the model parameters is extremely useful: gradients allow one to optimize the model efficiently; gradients allow the computation of the information matrix to forecast the precision with which parameters may be estimated given a specific experimental design; and gradients enable the use of a wider range of optimization and posterior sampling methods that perform well in high-dimensional parameter spaces, in particular, Hamiltonian Monte Carlo \citep[HMC; ][]{Duane1987,Neal2011,Betancourt2017,Monnahan2016,Tuchow2019}.

In addition, some previous models do not compute time-integrated exposures due to the computational cost of "super-sampling" the sky-plane coordinates with an \textit{N}-body model. This issue is compounded if one wishes to also compute derivatives. Currently, the models that are differentiable use approximate dynamical models to offset this cost, such as a Keplerian (eg. \texttt{exoplanet}) or by assuming a linear trajectory extrapolated from the position and velocity at the transit mid-point (eg. \texttt{jnkepler}). The former will not capture transit-timing variations, and the latter will miss any non-linearity in the trajectory during transit.

We present a fully analytically differentiable photodynamical model for fitting light curves of transiting multiplanet systems. Our method is a composite of three models: 1). the differentiable \textit{N}-body integrator, \nbg\footnote{\url{https://github.com/ericagol/NbodyGradient.jl}}, from \citet[][hereafter AHL21]{AHL2021} combined with 2). a high-order series expansion from \citet[][hereafter PK20]{ PK2020} to efficiently compute the sky-positions, and 3). the differentiable transit model, \limbdark\footnote{\url{https://github.com/rodluger/Limbdark.jl}}, from \citet[][hereafter ALFM20]{Agol2020}. We allow for quadratic limb-darkening, which could be extended to polynomial limb-darkening models \citep{Mandel2002, Gimenez2006, Agol2020}; we compute both time-integrated and instantaneous exposures.\footnote{We do not account for `mutual events' (conjunction of three or more bodies during a transit, e.g.\ \citealt{Gordon2022}). In principle, this can be done by implementing the model from \citet{Pal2012}, and computing the derivatives with respect to the model parameters.} Finally, the code allows for an arbitrary hierarchy of bound orbits as initial conditions using the formalism specified in \citet{Hamers2016}.

We start by describing the photodynamical algorithm in \S \ref{sec:algorithm}, followed by a description of the propagation of the derivatives using the chain rule in \S \ref{sec:derivatives}. We then describe the implementation of the model in the Julia language in \S \ref{sec:implementation} and compare our implementation with existing open source codes. Finally, we demonstrate some of the benefits of having a differentiable model by computing a likelihood profile for a synthetic data set and carrying out Bayesian inference on a sequence of simulated data sets with different numbers of planets (\S \ref{sec:application}). An appendix (\S \ref{app:ics}) describes the computation of the Jacobian of the initial conditions going from Cartesian coordinates to orbital elements, which is implemented in the \texttt{NbodyGradient.jl} package. We have implemented our model in Julia under \texttt{Photodynamics.jl}\footnote{\url{https://github.com/langfzac/photodynamics.jl}}, which is open-source and available on GitHub.

\section{Outline of the Algorithm}\label{sec:algorithm}
We now outline the fully differentiable photodynamical model. At the highest level, this work is a composite of previously developed methods. The novelty of our approach is that this particular composition lends itself to high-performance, analytic derivatives. 

\begin{figure*}
    \centering
    \includegraphics[width=0.6\hsize]{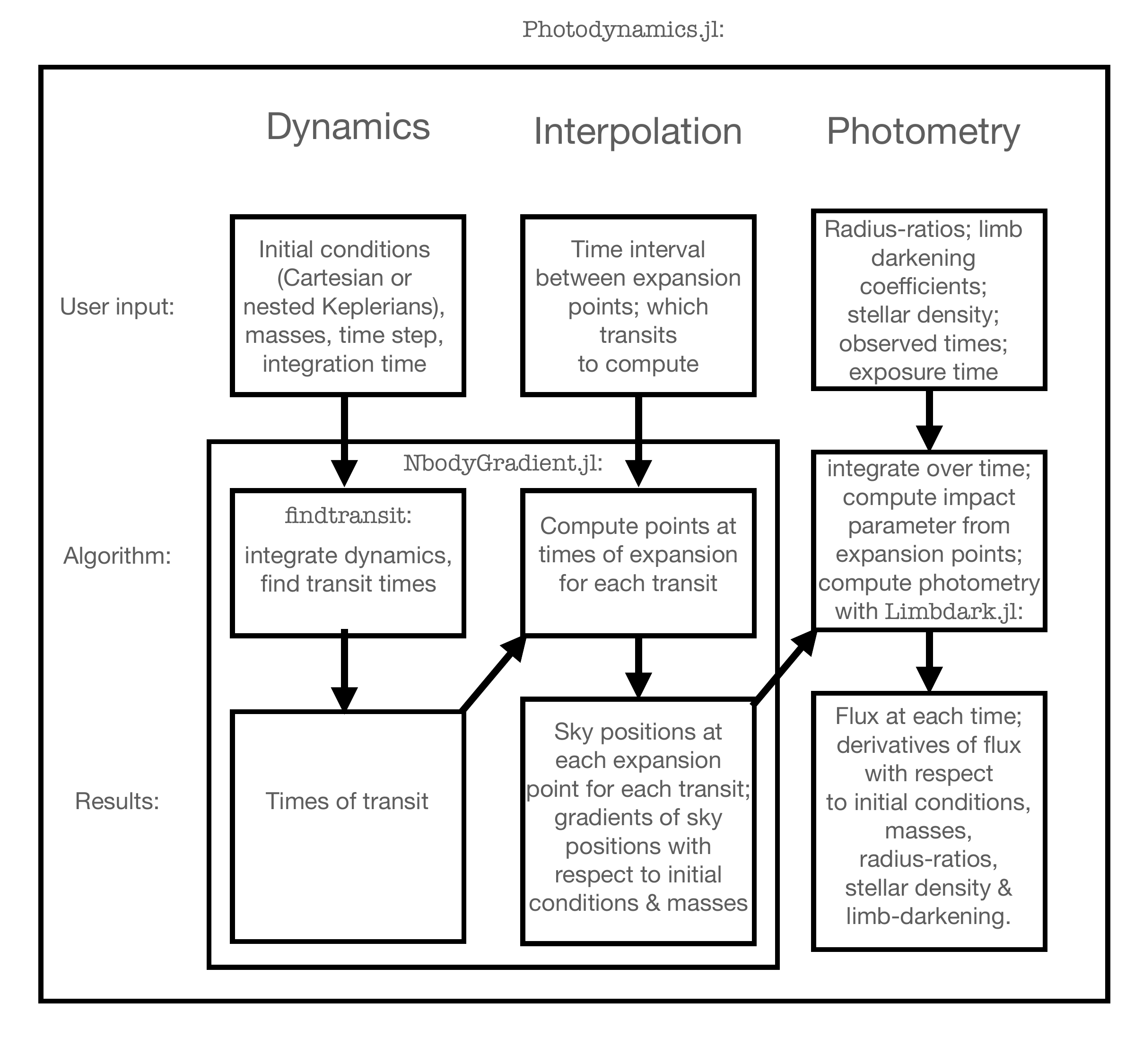}
    \caption{Flow chart of the photodynamical model.}
    \label{fig:code_flowchart}
\end{figure*}

The algorithm can be effectively carried out in 2 main steps: 1) the \textit{N}-body step, i.e. the dynamical model, and 2) the photometry step. In brief, we first use an \textit{N}-body model to find, and compute sky-plane positions about each transit. We then expand the sky-plane positions at each transit time following PK20 and use expansion-computed positions as input for the photometric model for each exposure time. Finally, we use the ALFM20 transit model to compute the flux as a function of the impact parameter at a given time, the planet-star radius ratio, and the quadratic limb-darkening coefficients. In addition, the model provides derivatives of the flux with respect to each of these parameters. We then compute the gradient of the flux with respect to the initial conditions of the \textit{N}-body model by propagating the derivatives of the impact parameter with respect to the initial coordinates and masses via the chain rule. Figure \ref{fig:code_flowchart} shows this model as a flow chart, and we detail the process in the following sections.

\subsection{\textit{N}-body Step}\label{ssec:nbody}
The computational cost is often a deterrent from implementing a photodynamical model. In a parameter inference context, the cost of evaluating the likelihood function will be dominated by computing the sky-plane position (i.e. impact parameter) around each transit, particularly if one is using full \textit{N}-body integration. To make the problem worse, we often wish to compute time-integrated exposures which requires repeated evaluations of the dynamical model for each exposure. This particular issue is addressed by PK20, where the authors describe a Taylor series expansion of the sky-plane coordinates about the transit mid-point using a 7-point finite difference method \citep{Fornberg1988}. By computing only 7 positions at each transit, the method provides an analytic function for the sky-plane position, which is accurate even for eccentric and short period systems. \cite{korth2020} describe a photodynamical model using this method, from which ours differs in choice of particular \textit{N}-body and transit models -- leading to an analytically differentiable model.

In order to obtain derivatives, we use the differentiable 4th-order sympletic AHL21 integrator to compute the expansion points, which are evenly spaced in time and centered at the given transit midpoint. AHL21 computes the Jacobian of the current Cartesian coordinates with respect to the initial Cartesian coordinates and masses. 

An example of pseudo-code for the \textit{N}-body step is given in Algorithm \ref{alg:transit-series}.

\begin{algorithm}
\KwData{Initial Cartesian coordinates and masses at time $t=t_0$}
\KwResult{Expansion points for computing impact parameter}
\For{$t-t_0 < t_\mathrm{tmax}$}{
    Take a step $h$ with ALH21 and compute transit times if they occur\;
    \If{any transits occurred} {
        \For{each transiting body} {
            Generate set of 7 evenly spaced ($dt$) expansion point times, centered on the transit mid-point\;
            \For{each expansion point time} {
                integrate to expansion point time\;
                record sky-plane positions (x,y) and derivatives\;
            }
        }
    }
}
\caption{\textit{N}-body Step: computing the PK20 expansion points using AHL21 integrator.}
\label{alg:transit-series}
\end{algorithm}

\subsection{Photometry Step}\label{ssec:photometry}
The photometric model assumes a spherically-symmetric, limb-darkened star (or source), which means that the only dependence of the photometry with respect to the dynamical integration is through the impact parameter of the planet (or other foreground body) with respect to the center of the source, $b(t)$. Rather than directly computing $b(t)$ from the dynamical integration, we use the PK20 quartic expansion for the sky-plane positions at each transit or eclipse. Once the expansion points are known (via the \textit{N}-body step \ref{ssec:nbody}), we can quickly and accurately compute the impact parameter (and its derivatives) at any time during transit. This enables adaptive integration of each time step without the need to re-integrate the dynamics to every instant in time. We discuss the accuracy of this approximation below.

Combined with the ALFM20 transit model, this enables efficient computations of the instantaneous flux and the gradient with respect to the dynamical parameters, radius ratio, and limb-darkening coefficients. As described in \citet{Agol2020}, we can also carry out time-integration with an adaptive Simpson integration, in which the desired precision and number of iteration levels are specified. Then, we can evaluate the derivatives at the locations found during the adaptation step to obtain the corresponding time-integrated derivatives.

Assuming no mutual events (e.g. a planet transiting both another planet and the star at the same time), we can realize a full light curve by summing the contributions from each transit for every exposure in the time-series. The derivative "light curves" are computed in the same way. We have provided the pseudo-code in Algorithm \ref{alg:lightcurve}. 

\begin{algorithm}
 \KwData{Expansion points and derivatives from \ref{alg:transit-series}, quadratic limb-darkening coefficients, star-planet radius ratios, radius of the star, and a set of times and exposure times for which to compute flux.}
    \KwResult{Flux and derivatives evaluated at the input times.}
    \For{each transit time}{
        \For{each exposure time} {
            \If{exposure is within transit}{
                Integrate ALFM20 model over exposure duration using PK20 to compute impact parameter and derivatives\;
                Add flux and derivatives to running total for current exposure time\;
            }
        }
        Normalize all fluxes and derivatives by exposure time\;
    }
    \caption{Photometry step: computing the flux and derivatives at each exposure time using ALFM20 transit model with the PK20 expansion as input.}
    \label{alg:lightcurve}
\end{algorithm}

\section{Derivatives of the Algorithm}\label{sec:derivatives}
Here, we discuss the derivatives of flux with respect to the model parameters. Of interest to us are the derivatives with respect to the \textit{N}-body initial conditions and the transit model parameters. For the transit model, only the derivative with respect to the impact parameter is modified by the dynamics. That is, ALFM20 provides the derivatives with respect to the transit parameters, and we only need to propagate the derivatives of the \textit{N}-body model to the final flux derivatives through the impact parameter. Since ALH21 computes the \textit{N}-body derivatives, we need only derive the intermediate partial derivatives of the impact parameter with respect to the PK20 expansion points. 

A quick note on notation: We denote vectors with bold, \textit{lower}-case letters, i.e. $\mathbf{x}$, and matrices with bold, \textit{upper}-case letters, i.e. $\mathbf{M}$. We define them to be column vectors, and therefore, the transpose, $\mathbf{x}^{\intercal}$, is a row vector. Then, given a scalar function of a vector, $f(\mathbf{x})$, we choose to define the gradient, $\der{f}{\mathbf{x}}$, as a row vector. Given a vector function of a scalar, $\mathbf{f}(x)$, the derivative, $\der{\mathbf{f}}{x}$, is a column vector. The Jacobian matrix then follows as $\mathbf{J}\left(\mathbf{f}(\mathbf{x})\right) = \left<\der{f_1}{\mathbf{x}}, \der{f_2}{\mathbf{x}}, ..., \der{f_N}{\mathbf{x}} \right>^{\intercal} = \der{\mathbf{f}}{\mathbf{x}}$.

\subsection{Derivatives of the dynamical model}
Our goal in this section is to derive the gradient of the impact parameter for a pair of bodies, $b(t)$, with respect to the initial Cartesian coordinates and masses, $\nbq(t_0) = \nbq_0$.
From \cite{PK2020}, the impact parameter is computed from the magnitude of the relative sky-plane position between the star and planet, $\mathbf{l}$, given by
\begin{eqnarray}\label{eqn:impact}
    \mathbf{l}(t) &=& \mathbf{l_0} + \mathbf{v}t + \frac{1}{2}\mathbf{a}t^2 + \frac{1}{6}\mathbf{j}t^3 + \frac{1}{24}\mathbf{s}t^4,\\
     &=& \left<l_x(t), l_y(t)\right>^{\intercal}, \nonumber \\
    b(t) &=& \frac{| \mathbf{l(t)} |}{R_*}.
\end{eqnarray}
Here, $\mathbf{v}$, $\mathbf{a}$, $\mathbf{j}$, $\mathbf{s}$ are the sequence of time-derivatives of $\mathbf{l}$, i.e.\ the 2-D sky velocity, acceleration, jerk, and snap vectors, and $R_*$ is the radius of the luminous body being transited. The time derivatives are approximated with finite differences \cite[see equations 2-5\footnote{We note that equation 5 in \cite{PK2020} has a typo: the coefficient of $d_3$ should be 1, not 2. Our code contains the correction.} in][]{PK2020}, based on the expansion coefficients computed in \citet{Fornberg1988}. The expressions are simply a function of the 7 expansion points, $\mathbf{x}$ and $\mathbf{y}$, which are computed from the \textit{N}-body integrator at the expansion times. The authors show that a time-step of 0.02 d between each expansion point is appropriate for most systems, including highly eccentric and short-period planets (see \citealt{PK2020} for details). Each time-derivative finite difference equation has a static set of coefficients. We define the matrix $\mathbf{C}$, where each row is the set of coefficients for each finite-difference time-derivative term. Then, the time-derivatives for the PK20 series expansion are computed approximately by
\begin{eqnarray}\label{eqn:components}
    \mathbf{c}_x = \mathbf{C}\mathbf{x},
\end{eqnarray}
where $\mathbf{c}_x$ is the vector of 1st- through 4th-order time-derivatives for the $x$ component of $\mathbf{l}$. The $y$ components are computed in the same way -- swapping $y$ for $x$ in the following and preceding equations. To represent $l_x$ with this formalism, we simply take the dot product: 
\begin{eqnarray}
    l_x = \mathbf{c}_x \cdot \mathbf{t},
\end{eqnarray}
where $\mathbf{t} = \left<1, t, t^2, t^3, t^4\right>^{\intercal}$, and 
\begin{eqnarray}
    b = \frac{\sqrt{l_x^2 + l_y^2}}{R_*}.
\end{eqnarray}
The gradient of $b$ with respect to $\nbq_0$\;is then
\begin{eqnarray}
\der{b}{\nbq_0} = \der{b}{l_x}\der{l_x}{\mathbf{c}_x}\der{\mathbf{c}_x}{\mathbf{x}}\der{\mathbf{x}}{\nbq_0} + \der{b}{l_y}\der{l_y}{\mathbf{c}_y}\der{\mathbf{c}_y}{\mathbf{y}}\der{\mathbf{y}}{\nbq_0},
\end{eqnarray}
where the intermediate derivatives are
\begin{eqnarray}
    \der{b}{l_x} &=& \frac{l_x}{\sqrt{l_x^2 + l_y^2}} \frac{1}{R_*}= \frac{l_x}{bR_*^2}, \label{eqn:dbdl} \\
    \der{l_x}{\mathbf{c}_x} &=& \mathbf{t}^{\intercal}, \\ 
    \der{\mathbf{c}_x}{\mathbf{x}} &=& \mathbf{C},
\end{eqnarray}
and $\der{\mathbf{x}}{\nbq_0}$ is the Jacobian of the seven $x$ components of the PK20 expansion points with respect to the initial coordinates and masses computed by the AHL21 integration\footnote{Note that equation \ref{eqn:dbdl} appears to be singular when $b \rightarrow 0$, which can cause numerical issues. Although, we find this case to be extremely unlikely in practice.}. We can then simplify to the final expression for the gradient of the impact parameter: 
\begin{eqnarray}
\der{b}{\nbq_0} &=& \frac{l_x}{bR_*^2} \mathbf{t}^{\intercal} \mathbf{C} \der{\mathbf{x}}{\nbq_0} + \frac{l_y}{bR_*^2} \mathbf{t}^{\intercal} \mathbf{C} \der{\mathbf{y}}{\nbq_0}. \nonumber \\
 &=& \frac{\mathbf{t}^{\intercal} \mathbf{C}}{bR_*^2}\left(l_x  \der{\mathbf{x}}{\nbq_0} + l_y \der{\mathbf{y}}{\nbq_0}\right). \label{eqn:dbdq}
\end{eqnarray}

We note the impact parameter also depends on the stellar radius, $R_*$, but this has no effect on the dynamical derivatives. However, for completeness, the derivative of b with respect to the stellar radius is
\begin{eqnarray}
    \der{b}{R_*} = -\frac{| \mathbf{l} |}{R_*^2} = -\frac{b}{R_*}.
\end{eqnarray}

\subsection{Derivatives of the flux}
With the derivatives of the impact parameter computed for an instant in time, we complete the computation of the flux, and the derivatives of the flux with respect to the initial conditions.

For a pair of bodies undergoing transit, with radii $R_p$ (planet) and $R_*$ (star), the limb-darkened transit flux model can be expressed as $F(k,b(t),\left\{u_i\right\})$, where $k=R_p/R_*$, $b(t)$ we have introduced above, and $\left\{u_i\right\}$ are the limb-darkening parameters of the star, or other luminous body \citep{Agol2020}. Thus, the dependence upon the initial orbital elements and masses arises from the impact parameter, $b(t)$, discussed in the last section. 

The derivatives of $F$ with respect to each of the three input parameters (e.g.\ $\partial F/\partial b$) are given in \cite{Agol2020}.\footnote{Note that in \cite{Agol2020} the variable $r$ is used instead of $k$ to represent the radius-ratio.} Using the chain rule, the expression for the gradient of the flux with respect to the initial Cartesian coordinates and masses of the system are simply:
\begin{eqnarray}
    \der{F}{\nbq_0} = \der{F}{b}\der{b}{\nbq_0}.
\end{eqnarray}
As discussed above, the derivatives need to be integrated over each exposure along with the flux to obtain the exposure-time averaged flux and its derivatives.

If one is using orbital elements (e.g. \ref{app:orbital_elements}) to specify initial conditions, then the Jacobian matrix of the transformation, $\der{\nbq_0}{\nbeta}$, may be applied to obtain the gradient with respect to an arbitrary set of initial orbital elements, $\nbeta$. That is,
\begin{eqnarray}
    \der{F}{\nbeta} = \der{F}{\nbq_0}\der{\nbq_0}{\nbeta}.
\end{eqnarray}

The derivatives of flux with respect to $k$ and $\left\{u_i\right\}$ are given in \citet{Agol2020}, which may be transformed to derivatives with respect to $R_p$ and $R_*$ with another step, i.e.,
\begin{eqnarray}
\frac{\partial F}{\partial R_p} &=& \frac{\partial F}{\partial k}\frac{1}{R_*},\label{eq:dfdrp}\\
\frac{\partial F}{\partial R_*} &=& -\frac{\partial F}{\partial b}\frac{b}{R_*}.\label{eq:dfdrs}
\end{eqnarray}
Even though the flux depends on $R_*$ through $k$, we can neglect the partial derivative of the flux with respect to the $k$ term in Equation \ref{eq:dfdrs} since $k$ is held fixed when varying $R_*$ (similarly for Equation \ref{eq:dfdrp} since $R_*$ is fixed).  This completes our description of the computation of the flux and its derivatives with respect to every model parameter.  We turn next to the coding up of this algorithm and its performance.

\section{Implementation and Performance}\label{sec:implementation}
We now discuss an implementation of the photodynamical model outlined in the preceding sections. We've written a \texttt{Julia} \citep{Julia} package that is open-source and publicly available on GitHub as \pdjl\footnote{\url{https://github.com/langfzac/Photodynamics.jl}}. In the following sections, we look at details of the implementation and we compare \pdjl to the previously developed photodynamics code \texttt{photodynam}\footnote{\url{https://github.com/dfm/photodynam}} \citep{Carter2011} in \S\ref{ssec:comparison}.

\subsection{Units, Coordinates, Conventions, and Initial Conditions}
\pdjl uses masses in solar masses $M_\odot$, time in days, and distance in AU. The right-handed coordinate system is set up so that the positive z-axis is pointing away from the observer, and the positive x-axis points to the right along the horizontal. See appendix \ref{app:ics} for details.

In practice, photodynamics is insensitive to the absolute radii or masses of the bodies in a system. The reason is that the depth of a transit solely depends on the radius-ratios \citep{Mandel2002}, while the transit-timing variations depend upon the mass-ratios \citep{AgolFabrycky2018}. In contrast, the stellar density {\it can} be constrained with a photodynamical model as transit durations decrease as $\rho_*^{-1/3}$, holding all other parameters fixed \citep{Seager2003}. Thus, we need to specify the stellar density to compute a photodynamical model. 

When combined with other techniques or other constraints, such as astrometry or spectroscopy, the stellar radius and mass can be derived, in principle. Hence, we let the stellar radius and mass be free parameters, but allowing only one to vary suffices to allow the stellar density to vary. For the planets, we specify the masses, and radius ratios, which scale as the square root of the transit depth \citep[the maximum percent change in flux during the transit;][]{Mandel2002, Seager2003}.

To finish specifying the transit model, we need to give the star a limb-darkening model. The \limbdark package enables polynomial limb-darkening to be used. The current implementation of \pdjl uses quadratic limbdarkening -- parameterized by $u_1$ and $u_2$.

As for the \textit{N}-body initial conditions, we specify either the initial Cartesian coordinates (\S\ref{app:cartesian_coordinates} or \citealt{Agol2021}), or the initial nested orbital elements (\S\ref{app:orbital_elements}) following \cite{Hamers2016}. In principle, an arbitrary hierarchy of Keplerians may be used to specify the initial conditions; however, our initial implementation only contains nested Keplerians, as needed to describe a multiplanet system or a bound hierarchical stellar system. We choose our set of orbital elements to be: the mass of the star $M_*$ and the mass $M_i$, period $P_i$, time of initial transit $t_{0,i}$, eccentricity $e_i$, argument of periastron $\omega_i$, inclination $I_i$, and longitude of ascending node $\Omega_i$ of each of the $N$ planets with respect to the center-of-mass of the inner bodies, where $i=1,2,3,...,N$. We then transform to Cartesian coordinates for integration, and compute the derivatives of the transformation in order to obtain the derivatives of the Cartesian coordinates at some later time with respect to the initial orbital elements.

The set of free parameters for a system with one star and $N$ planets is given in Table \ref{tab:params}. However, the exact parameterization may change depending on the context, but we list parameters that are used directly by the code and that are generally of interest. One can also use Cartesian coordinates to initialize the \textit{N}-body integrator.

\begin{table}
\caption{Free parameters for a system with 1 star and N planets. The $i$ subscript represents the $i$-th planet in the system.}
\begin{tabular}{ll}
\hline
Symbol & Description \\ \hline
$M_*$                      & Stellar mass [M$_\odot$]                     \\
$M_i$                      & Planet mass [M$_\odot$]                      \\
$P_i$                      & Period [Days]                                \\
$t_{0,i}$                  & Time of initial transit [Days]               \\
$e_i$                      & Eccentricity                                 \\
$\omega_i$                 & Argument of periastron [rad]                 \\
$I_i$                      & Inclination [rad]                            \\
$\Omega_i$                 & Longitude of Ascending Node [rad]            \\
$k_i$                      & Planet-star radius ratio                                 \\
$u_n$                      & Quadratic limbdarkening coefficients ($n=1,2$) \\
$R_*$                      & Stellar radius [AU]                         \\
\hline
\end{tabular}
\label{tab:params}
\end{table}

\begin{figure*}
\centering
\includegraphics[width=0.8\textwidth]{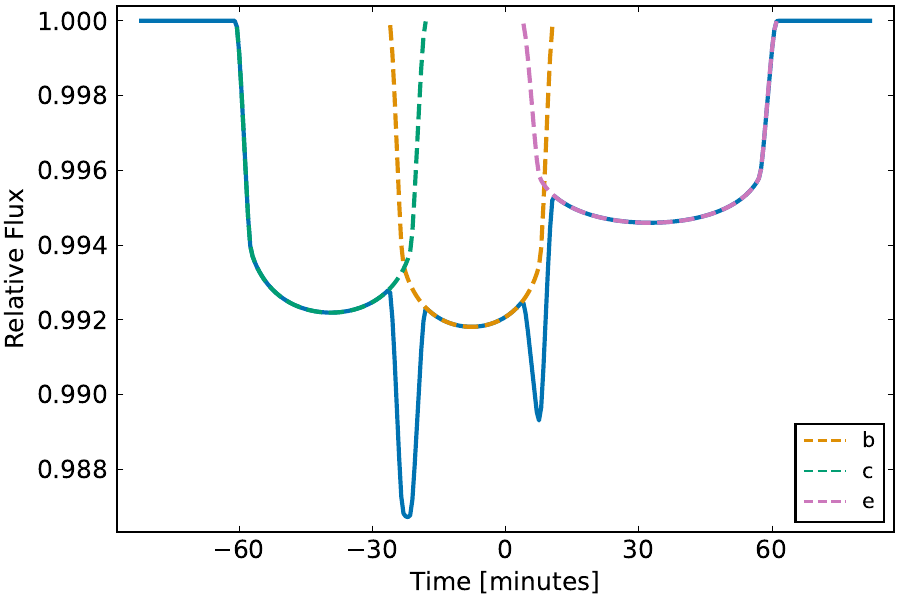}
\caption{Relative flux versus time in minutes for 3 planet (b,c,e) transits. The solid curve shows how the observed light curve would appear, and the dashed lines are the individual contributions for each transiting planet.}
\label{fig:transit_triple}
\end{figure*}

\begin{figure*}
\centering
\includegraphics[width=0.95\textwidth]{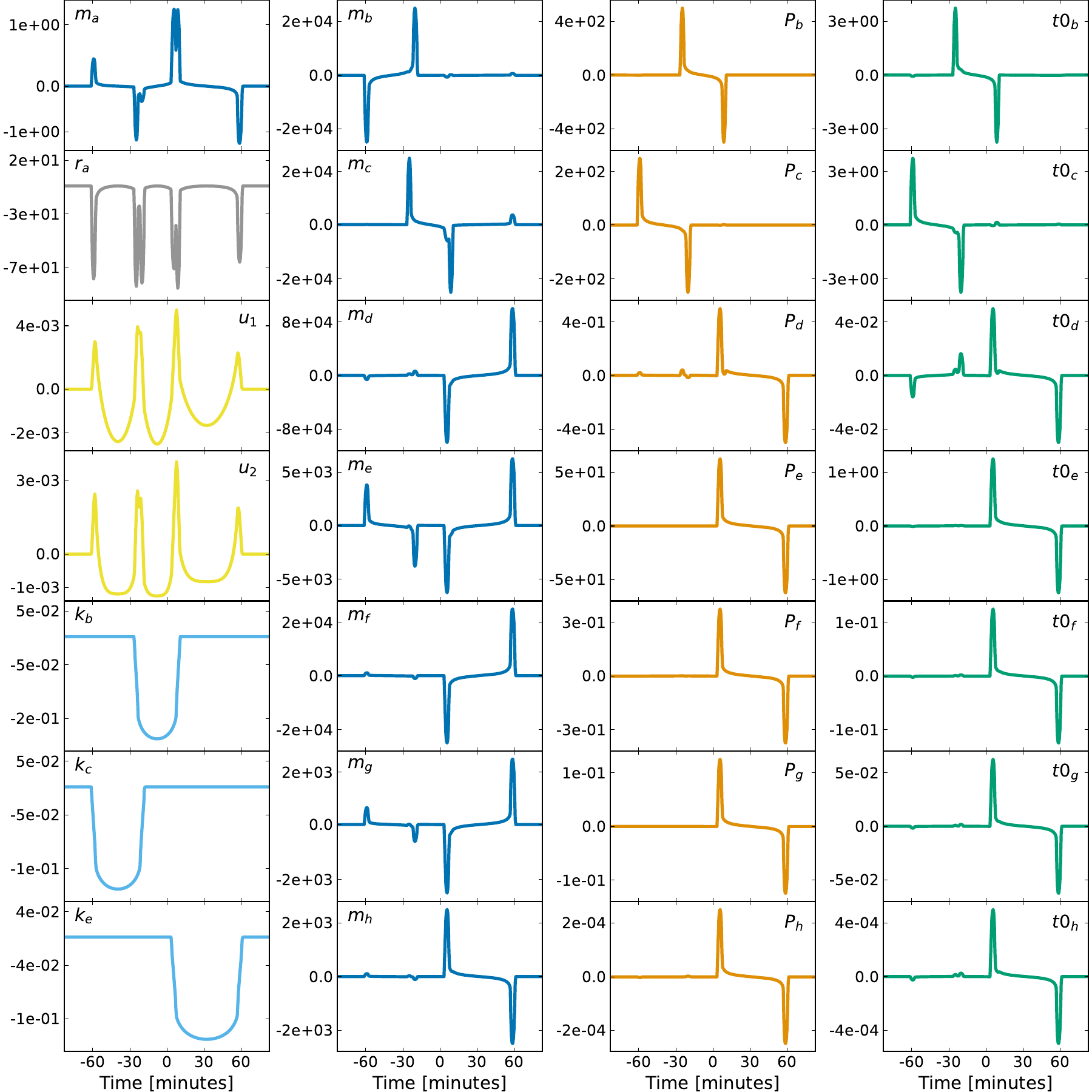}
\caption{Derivatives of the flux with respect to the labeled model parameter versus time in minutes for the 3 transits in Figure \ref{fig:transit_triple}. The times are with respect to an arbitrary mid point. The first column shows the stellar mass, stellar radius, both limbdarkening coefficients, and the 3 relevant radius ratios. The next 3 columns are the masses, periods, and initial times of transit for each planet (b-h), respectively.}
\label{fig:derivatives1}
\end{figure*}

\begin{figure*}
\centering
\includegraphics[width=0.95\textwidth]{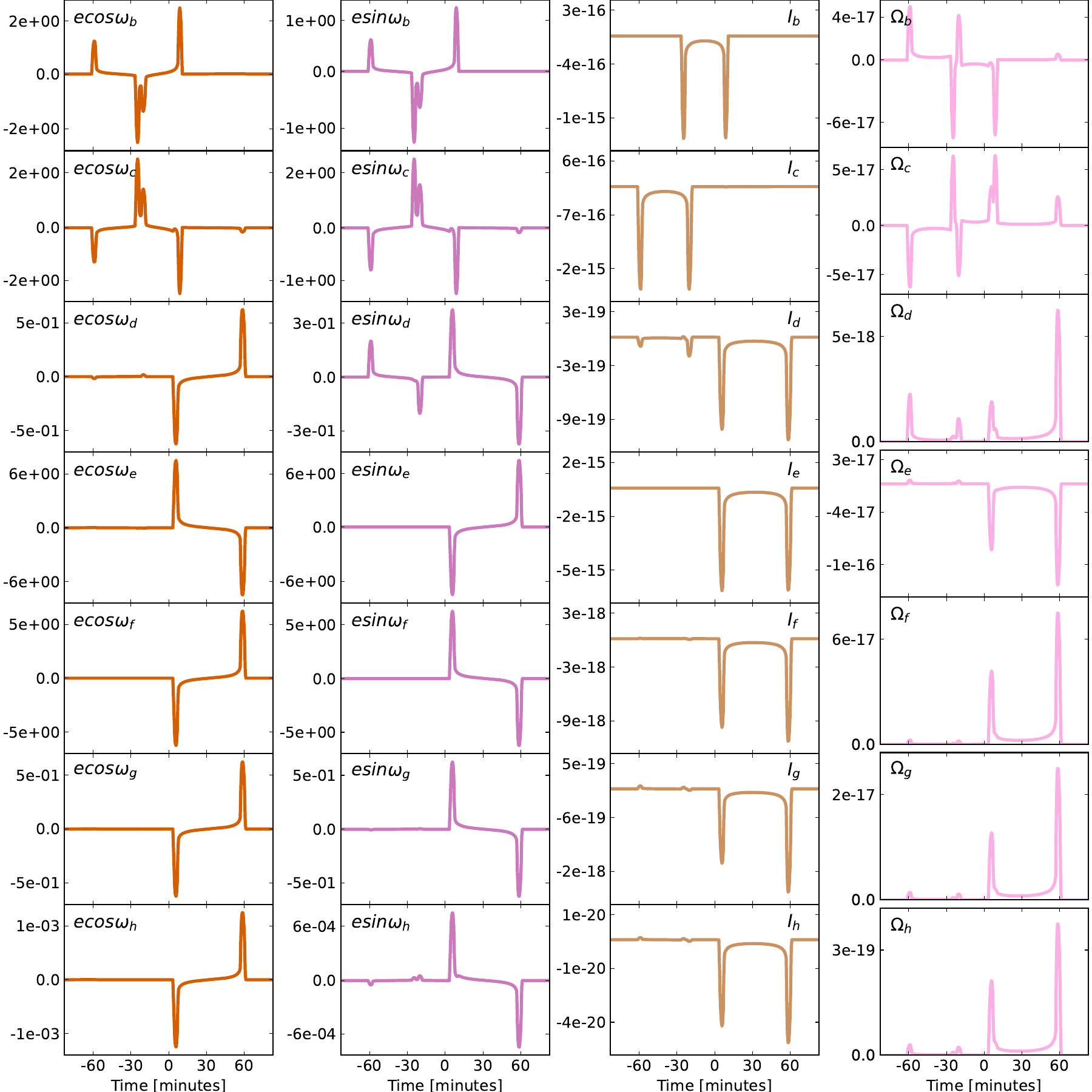}
\caption{Derivatives of the flux with respect to the labeled model parameter versus time in minutes for the 3 transits in Figure \ref{fig:transit_triple}. The times are with respect to an arbitrary mid point. The columns are the eccentricity vector components, the inclination, and the longitude of ascending node, respectively, for planets b-h.}
\label{fig:derivatives2}
\end{figure*}

\subsection{Implementation in Julia}
At the highest level, \pdjl is a composite of existing publicly available software: \nbg for the \textit{N}-body integration and \limbdark for the transit model. We extend \nbg to include computation of the PK20 expansion points, and to propagate derivatives of the coordinates with respect to the initial conditions\footnote{To be specific, the PK20 expansion code has been implemented in \pdjl, and the initial conditions for the orbital elements and masses have been implemented directly into \nbg.}. We wrote our own version of the PK20 expansion method, which also computes the derivatives, and re-implemented the version of Simpson's rule \citep{kuncir} from ALFM20 for the time-integration.

Using Julia comes with a number of advantages: just-in-time compilation allows Julia code to compete with \texttt{C} programs in performance; Multiple-dispatch allows us to easily run the code at different floating-point precision and add features without appreciably changing the user interface; Code can be run interactively via the built-in REPL, Jupyter notebooks \citep{jupyter} Pluto notebooks \citep{pluto}, or as scripts in a High-Performance Computing context; our analytic gradients can be hooked up to any of the various automatic differentiation (auto-diff, e.g., see \texttt{JuliaDiff}\footnote{\url{https://github.com/JuliaDiff/}}), which enables use of the robust Bayesian inference ecosystem (ie. \texttt{Turing.jl} \citep{turing}, \texttt{DynamicHMC.jl} \citep{dynamichmc}, etc.)\footnote{Currently, many of the tools we wish to use for Bayesian inference rely on automatic differentiation (auto-diff) for computing the gradient of the posterior probability. Since the derivatives of our model are straightforward to compute analytically (assuming the work from ALFM20 and AHL21 has already been done, of course), we choose to implement them instead of using auto-diff, which does impose limits on how the code may be structured. However, it is not always straightforward to simply supply analytic gradients to a posterior sampling code -- these often assume the user is using auto-diff. To get around this, one can implement an auto-diff "rule" for the model that simply supplies the the analytic derivatives from the algorithm one is using.}. 

We constructed the code with the case of repeated model evaluations (e.g. MCMC) in mind. While the \textit{N}-body integration dominates the computation time, we can still be deliberate about the code architecture to optimize performance. For this code, the main optimizations come from memory management. Julia uses a garbage collector and, by default, allocates mutable types (i.e. arrays) on the heap, so pre-allocating large arrays can speed up compute time. We pre-allocate any arrays for which the size depends on the particular set of model parameters, most of which also depend on the number of data points we wish to model. For example, the size of the array that holds the derivatives of the $x$ and $y$ coordinates of the PK20 expansion points scales as the product of the number of bodies and the number of total transits. We also make use of "statically sized arrays" or "static arrays" (\texttt{StaticArrays.jl}\footnote{\url{https://github.com/JuliaArrays/StaticArrays.jl}}) for any small, intermediate arrays that never change their size, such as the coefficients of the finite difference derivatives in the PK20 expansion. Using static arrays also allows the Julia compiler to optimize array operations (e.g. the dot-product) to account for the array's size\footnote{The advantage of using static arrays diminishes if the arrays become too large, which is why we do not use them for the entire code.}. 

We've written a comprehensive testing suite, which makes use of the continuous integration tools available to GitHub repositories. The tests cover 1) the accuracy of the analytic derivatives compared to \texttt{BigFloat} (256-bit float) precision central finite-difference derivatives at multiple points in the algorithm, 2) the accuracy and precision of the impact parameter computed with the PK20 expansion versus direct \textit{N}-body integration, and 3) the various intermediate and utility methods such as the point-of-contact computation, the accuracy of the Simpson's integration used for the time-integration, and the user-interface functionality.

\subsubsection{Derivative light curves}
Figure \ref{fig:transit_triple} shows a super-sampled (30 s, time-integrated exposures) light curve for the transits of 3 planets (analogues of TRAPPIST-1 b, c, and e) computed by \pdjl. We simulate a TRAPPIST-1-like, 7 planet configuration using similar initial conditions to those from \cite{Agol2021}. We plot the relative flux versus time in minutes from an arbitrary midpoint (chosen purely for display purposes). The solid line shows the computed light curve, and the dashed lines represent the individual transits of each planet.

Figures \ref{fig:derivatives1} and \ref{fig:derivatives2} display the "derivative light curves" for the transits in Figure \ref{fig:transit_triple}. Each plot is of the derivative of the flux with respect to the labeled model parameter versus time. We note that the y-axes are not the same, and that we are interested in showing the shape of the derivatives and the relative magnitude across similar parameters (eg. the contribution of each planet's mass). As expected, changes to the stellar mass, radius, and limb-darkening coefficients affect the transits of all three planets, and the radius ratios only affect the exposures during the transit of the corresponding planet. Other behavior present is not as straightforward to interpret, as changes to a single orbital element can significantly change the initial Cartesian coordinates of the system (see Appendix \ref{app:ics}). However, it is expected that changes to, say, the initial period of a planet's orbit would affect the exposures during ingress/egress -- when the planet begins and ends its occultation -- more than those around the midpoint of the transit. Figure \ref{fig:derivatives1} also demonstrates some interesting aspects of photodynamical models.  Even though transits of planets $d$ and $f$-$h$ do not appear in this light curve, the photodynamical model depends on parameters of each of these planets thanks to their dynamical impact on planets that are transiting. Next, this figure demonstrates the general point that the TTVs of a planet do not strongly depend on its own mass. The panel for the derivative of flux with respect to $m_b$ (first row, second column) shows that the transit of planet $b$ is only weakly dependent upon its own mass;  this is due to the fact that its acceleration of a body does not depend on its mass, but only on the masses of all of the other bodies in the system.  Since planet $b$ strongly perturbs the orbit of the adjacent planet $c$ thanks to growth of the inverse square gravity with proximity, the transit of planet $c$ has a derivative which depends strongly on the mass of planet $b$. The transit of planet $e$ only weakly depends on the mass of planet $b$ due to their remoteness from one another.  Figure \ref{fig:derivatives2} also illustrates an interesting aspect of transit-duration variations.  The 3rd column shows derivatives with respect to the inclinations of all of the planets in the system.  Mutually inclined planets will precess, causing their durations to change more strongly than the times of transit change. This is apparent in the derivatives, which look fairly symmetric for each transit (an inverted ``batman" curve), and most strongly depend upon the inclination of the transiting planet, and more weakly on the inclinations of more distant planets.

\subsubsection{A note on time-integration}
\cite{Agol2020} found that adaptive time-integration of the light curve works best if the exposures are split at each point of contact. We compute the times of the points of contact, $\{t_c\}$, for a pair of bodies numerically using the \texttt{Roots.jl} package (Newton's method) with the impact parameter computed from the series expansion. That is, we solve $b(t_c) = 1 \pm k$ where $b$ and $k$ are the impact parameter and radius of the occulter in units of the radius of the luminous body (where $+$ is for the first and fourth points of contact, and $-$ for the second and third, which only occur for a non-grazing transit). The root finder is initialized for each case with the time of transit plus or minus the series expansion time step. Then, if a contact point time falls within an exposure, we split the exposure in two -- integrating the two sub-exposures and taking the time-average. The result is a numerically-accurate time-integrated light curve.

\subsection{Comparison to high-precision numerical model}

The standard precision for computation with modern CPUs is double-precision (64-bit).  In this section, we test the numerical precision of the double-precision model by comparing light curves computed using double-precision with those computed in octuple-precision (64-bit versus\ 256-bit).  These are implemented as the \texttt{Float64} and \texttt{BigFloat} data types in Julia, respectively. Note that thanks to Julia's facility with multiple-dispatch, we can simply call the functions in \texttt{Photodynamics.jl} with input of the initial parameters in \texttt{BigFloat} precision, and the just-in-time compiler automatically recompiles to operate in \texttt{BigFloat}; i.e.\ the entire computation is carried out with 256-bit octuple precision. Also note that truncation and rounding errors in octuple precision should be greatly reduced compared with double-precision, so that this comparison allows us to diagnose the accumulation of numerical errors for the double-precision computation. 

\begin{figure*}
\centering
\includegraphics[width=0.8\textwidth]{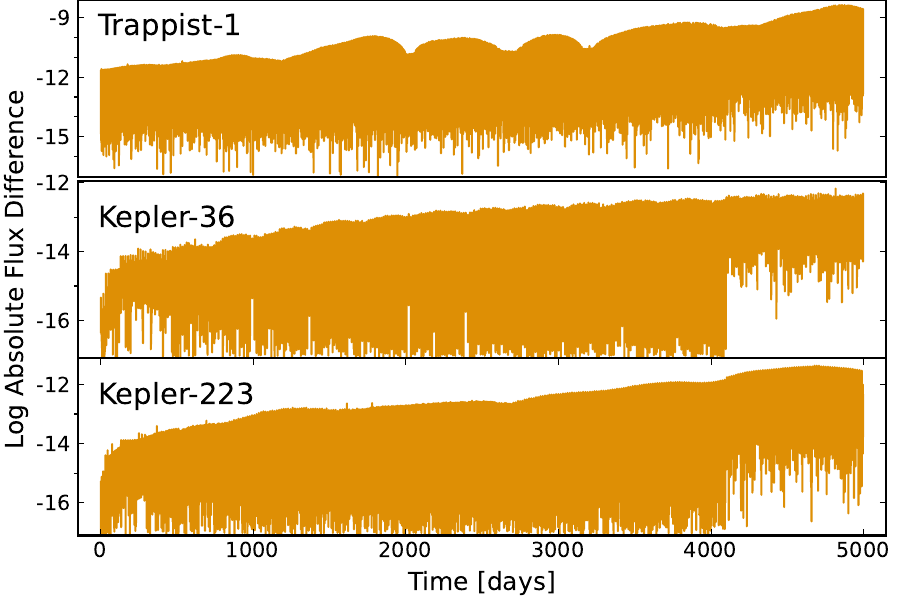}
\caption{The log of the absolute flux versus integration time for 3, multiplanet light curves.  Top: Trappist-1 -- 7 planets. Middle: Kepler-36 -- 2 planets. Bottom: Kepler-223 -- 4 planets. Over the length of each 5000 d simulation, the flux difference remains orders of magnitude lower than expected state of the art instrumental uncertainties.}
\label{fig:bigfloat}
\end{figure*}

We keep the initial conditions and other parameters identical for the comparison by simply converting the double-precision parameters into octuple-precision. Figure \ref{fig:bigfloat} shows the log of the absolute flux difference between \texttt{Float64} and \texttt{BigFloat}  simulations of light curves for three systems with multitransiting planets over 5000 d. The top panel shows TRAPPIST-1 \citep{Agol2021}, a 7-planet system; the middle panel shows Kepler-36 \citep{Carter2012}, a 2-planet system with large (hours) TTVs; the bottom panel shows Kepler-223 \citep{Mills2016}, a system with 4 planets in a resonant chain. The initial parameters were drawn from the system parameters reported in each paper. Each system is integrated for 5000 d with a timestep 1/40-th of the shortest orbital period for the system, and 2 min cadence photometry. While the numerical errors grow over the simulation, which is expected, we see that they remain well below the expected state-of-the-art instrument uncertainties (e.g. JWST at $\sim$10 parts per million \citep{Rustamkulov2022}) over a large time baseline.

\subsection{Comparison to photodynam} \label{ssec:comparison}

In this section we compare \pdjl with \texttt{photodynam}\footnote{\url{https://github.com/dfm/photodynam}} -- a photodynamics code written in \texttt{C}, which has been used in a number of light curve analyses \citep[e.g.][]{Carter2011,Carter2012}. This is not a true one-to-one comparision, as \photodynam uses distinct photometric model and integration schemes (see Table \ref{tab:codes}). However, the code can still serve as a benchmark, especially for the precision of the dynamical model. The \photodynam code uses the adaptive Bulirsch-Stoer algorithm \citep{numerical_recipes} to compute the sky-plane position at each time step. This allows for higher-precision integration at the cost of computation time compared to symplectic integrators like AHL21. \photodynam does not compute time-integrated exposures, so we compare to only the instantaneous flux variant of \pdjl. Both codes use quadratic limbdarkening, but \photodynam uses the path-integral method from \cite{Pal2012} to compute the flux.

\photodynam was originally used to compute light curves for eclipsing star systems, which require treatment of the light-travel time when computing relative positions of luminous bodies. For transiting exoplanets, this effect can often be ignored, so we turn off this feature in \photodynam for the purposes of comparison. Our comparison consists of using both codes to compute a light curve of a TRAPPIST-1 like system over 1600 d \citep[roughly the length of the data set used in][]{Agol2021} with 2 min cadence. For performance comparisons, we run \pdjl both with and without derivative computations and with and without integrated exposures. The two codes differ in initial condition specification, so to ensure that the \textit{N}-body integrations are consistent, we generate initial barycentric Cartesian coordinates by running \photodynam with a set of orbital elements. These Cartesian coordinates can then be used as initial conditions for \pdjl. For this comparison, we used the TRAPPIST-1 orbital elements from \cite{Agol2021}.

We choose the time step for each integrator to be the period of the innermost planet divided by 40. To compute the sky-positions for each exposure time, \photodynam uses an adaptive integration scheme, for which we set the tolerance to double float precision.

\begin{figure*}
\centering
\includegraphics[width=0.8\textwidth]{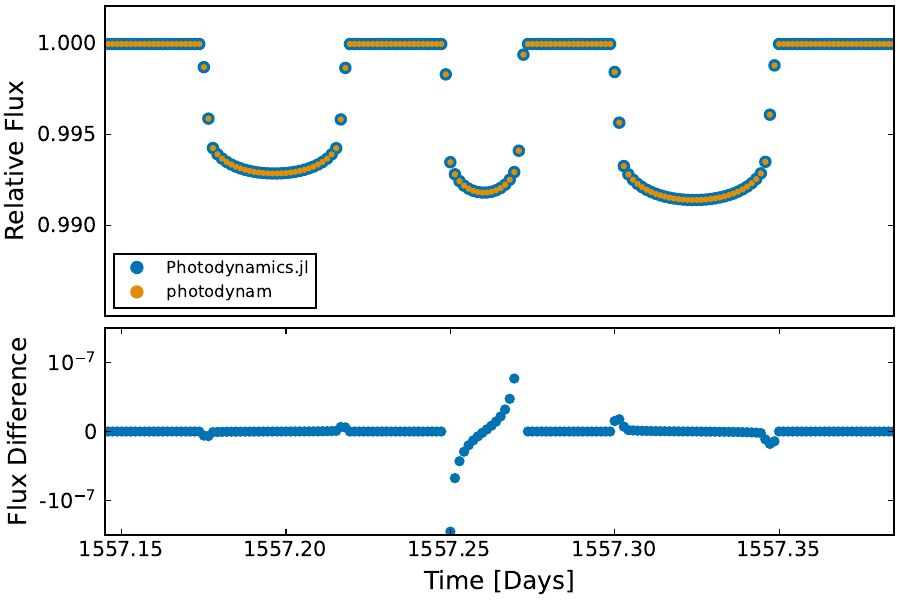}
\caption{A set of 3 transits computed by both \pdjl (blue; larger) and \photodynam (orange; smaller) (top), and the difference between them (bottom), versus time in days. These transits are near the end of the simulation, and demonstrate that the differences between the two algorithms remain well below the expected instrumental uncertainty for JWST.}
\label{fig:transit_single}
\end{figure*}

\begin{figure*}
\centering
\includegraphics[width=0.8\textwidth]{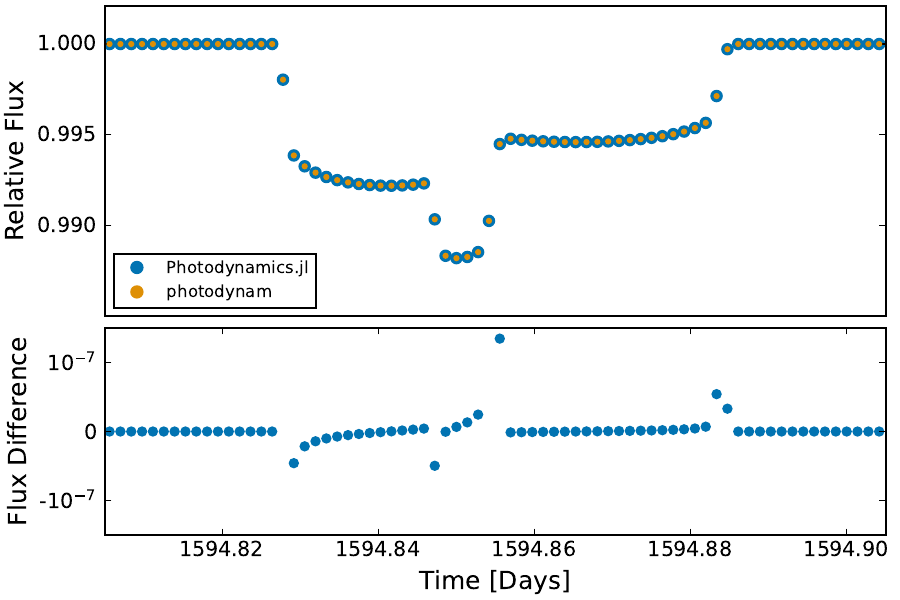}
\caption{The computed flux from both \pdjl (blue; larger) and \photodynam (orange; smaller) (top), and the difference between them (bottom), versus time in days for the simultaneous transit of 2 planets. As with the figure above, this event is near the end of the simulation.}
\label{fig:transit_double}
\end{figure*}

Figure \ref{fig:transit_single} shows a section of the light curve containing three individual transits computed by both codes (top) and the difference between them (bottom). The transits occur starting at roughly 1557 d, which is near the end of the 1600 d simulation, and show that the maximum error remains well below the expected noise floor of, for example, $\sim$10 ppm for JWST photometry \citep[e.g.][]{Schlawin2021}. Figure \ref{fig:transit_double} shows the transit of two planets simultaneously (top) and the residuals (bottom). We note that in all cases the largest errors are found during ingress and egress of the transit. This is expected as these exposures are much more sensitive to variations in the positions as opposed to when the planet is fully occulting the star.

\begin{figure}
\centering
\includegraphics[width=1.\columnwidth]{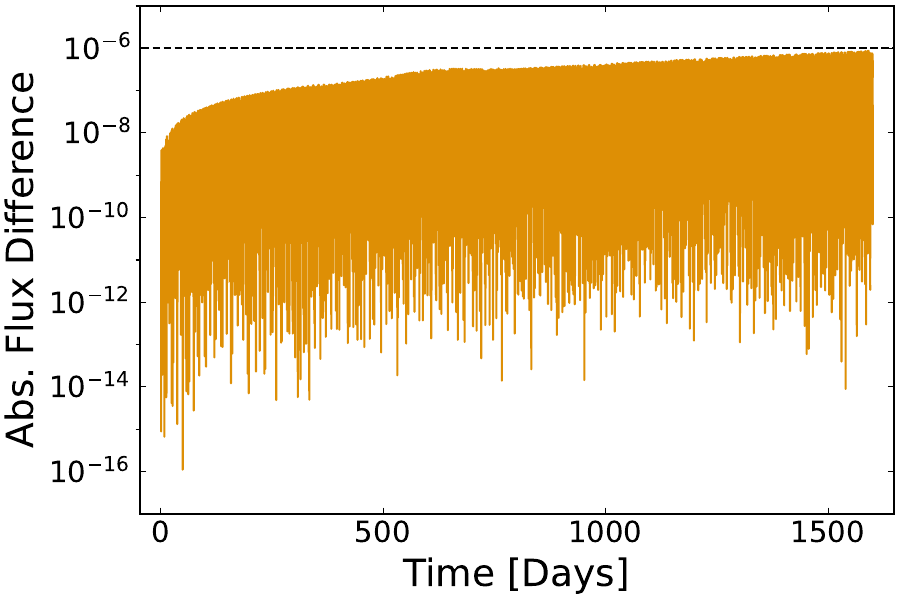}
\caption{The absolute value of the difference in flux between \pdjl and \photodynam for a 7 planet system over 1600 d with 2 min cadence. The black, dashed line represents a flux difference of $10^{-6}$.}
\label{fig:residuals}
\end{figure}

Figure \ref{fig:residuals} shows the difference between the codes for each exposure over the entire 1600 d of the simulation. The errors grow over time, which is expected due to the difference in integration schemes, leading to variations in the positions. Even with this growth, the errors remain below 1 ppm.\footnote{We noticed a bug in \photodynam, where a few transits had erroneous flux values. That is, some exposures were equal to 0 during the transit, with the rest of the exposures appearing as expected. This occurs rarely, so we remove these values from our comparison.}.

Clearly, \pdjl compares well in precision to \photodynam, despite using an ostensibly less precise dynamical model. We note that codes like \photodynam that use the \cite{Pal2012} method are able to compute mutual events, while our method cannot. However, these events are rare -- only 20 out of 2764 transits in this comparison were mutual events, and we assumed exactly edge on orbits for a 7 planet system.

We also compare the computational performance of each algorithm. For \photodynam, a \texttt{C} program, we simply time the execution of the program in the command line, which includes reading/writing to files. \pdjl is timed similarly -- starting with reading in the initial conditions files and outputting the resulting light curve data. We carried out this comparison using the \texttt{BenchmarkTools.jl} \citep{benchmarktools} framework, on a machine with an AMD EPYC 7443 24-core processor. \photodynam takes about 43.1 s to run and the equivalent variant of \pdjl takes about 4.9 s. Table \ref{tab:performance} shows the results of running each variant of \pdjl -- with/without derivatives and integrated exposures. As implemented, all cases of \pdjl out perform \photodynam.

We do want to note that there are discrepancies in implementation that, when accounted for, may change the performance difference between \pdjl and \photodynam. Mainly, as implemented, \photodynam writes the output to a file at every exposure time, while \pdjl saves intermediate results in memory. Writing to files is a known cause of overhead and thus could be a non-negligible contribution to the run time of \photodynam.

\begin{table}
\caption{Execution times for the comparison simulations described in Section \ref{ssec:comparison}. The codes simulate the light curve for a TRAPPIST-1 like, 7 planet system for 100 d with 2 min cadence. Each of the above cases uses the same initial conditions and other model parameters. For the \pdjl runs, the differences are simply whether we choose to compute derivatives and/or time-integrated exposures.}
\begin{tabular}{lcc}
\hline
Code:                          & \pdjl & \photodynam \\ \hline
non-integrated, w/o derivatives & 4.9 s          & 43.1 s    \\
non-integrated, w/ derivatives  & 5.5 s          & --          \\
integrated, w/o derivatives     & 29.2 s          & --          \\
integrated, w/ derivatives      & 31.3 s          & --         \\
\hline
\end{tabular}%
\label{tab:performance}
\end{table}

\section{Example Applications}\label{sec:application}

In this section, we demonstrate the benefits of a differentiable model in the context of parameter inference. We create a synthetic data set ($\S$ \ref{ssec:data}), adding white noise, and then optimize a fit to this simulated data set using the Levenberg-Marquardt method ($\S$ \ref{ssec:optim}). This is followed by a computation of the likelihood profiles of the model parameters ($\S$ \ref{ssec:likelihood-profile}). Finally, we carry out a comparison of posterior probability inference with Markov chain Monte Carlo using an affine-invariant sampler and using a no-u-turn sampler (NUTS) ($\S$ \ref{ssec:mcmc}).

\subsection{Optimization and Profile Likelihood}
\subsubsection{Synthetic Data set}\label{ssec:data}
We created a synthetic data set for a two-planet system which is in close proximity to a 4:3 resonance (periods of $\approx 6.626$ and $\approx 8.967$ d).  We chose an integration cadence of two minutes for a duration of 299 d to cover twice the TTV super-period.  The planets' orbits are coplanar with edge-on orbits.  The eccentricity vectors were chosen from a Normal distribution with a standard deviation of 0.01. We fixed the stellar mass and radius to 0.5 $R_\odot$ and $0.5 M_\odot$, we randomly chose quadratic limb-darkening coefficients. The planets' masses and radii had values in the ``super-puff" range, $\approx 3 M_\oplus$ and $8 M_\oplus$, and $\approx 5 R_\oplus$.

We chose a level of noise to allow for the detection of the TTVs with high significance.  The photometric uncertainty per exposure time was 300 parts per million.

\begin{figure*}
    \centering
    \includegraphics[width=0.95\linewidth]{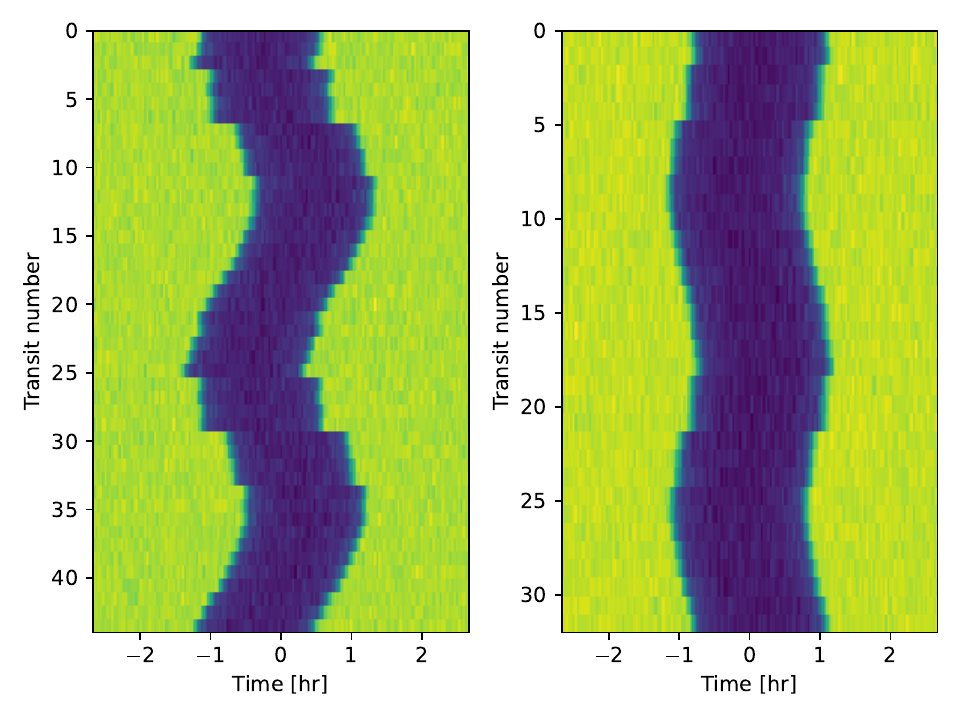}
    \caption{Riverplots of the synthetic photodynamical model used for testing the optimization with and without analytic derivatives.  Each row is a transit number plotted with a color scale in which the star is brighter (green) or dimmer (blue). Each pixel is a 2 min cadence, and the horizontal axis shows the time relative to the mean ephemeris of the transits of each planet. Left panel is for the inner planet and right panel for the outer.}
    \label{fig:riverplot}
\end{figure*}

The synthetic light curves are shown in Figure \ref{fig:riverplot} as a riverplot for each planet \citep{Carter2012}.  The ``river" meanders due to transit-timing variations due to proximity to the 4:3 resonance, and these are anti-correlated between the planets thanks to conservation of energy.  The inner (outer) planet has a smaller (larger) mass, and thus the TTVs of the outer (inner) planet are larger (smaller). 

\subsubsection{Optimization}\label{ssec:optim}
We carried out optimizations of the fit to the simulated data set using the Levenberg-Marquardt method as implemented in the Julia package \texttt{LsqFit.jl}. This method assumes a chi-square distribution for the negative log-likelihood. To reduce including portions of the data that are just noise, we used a window around each transit equal to 1/30 of the orbital period of the outer planet centered on each transit and discarded the data outside of these transit windows.

Before carrying out the optimization, we rescale each of the parameters so that their derivatives have similar magnitudes.  In practice this involved multiplying the periods by $10^6$, initial times of transit by $10^4$, eccentricities by $10^6$, mass-ratios by $10^9$, and radius-ratios by $10^3$. We found that this helped the Levenberg-Marquardt optimizer to converge more efficiently.

The first optimization was carried out without the use of analytic derivatives. The \texttt{LsqFit.jl} \texttt{curve\_fit} function defaults to numerically compute derivatives with finite differences to estimate the Jacobian.  We find that these numerically computed derivatives are not as accurate as our analytic derivatives, which has a negative impact on the convergence of the optimization algorithm. The algorithm does not converge to the global optimum when using numerically computed derivatives. Consequently, the standard errors returned from the algorithm are inaccurate.

The second approach uses the analytic derivatives provided by \texttt{Photodynamics.jl}. This optimization runs more quickly (about 10 times faster\footnote{We note that these types of performance differences can vary substantially with, for example, length of the data set or the shape of the likelihood near the optimum.}), and converges to the global optimum due to the better accuracy of the derivatives compared with the numerical finite difference case. This demonstrates the significant advantage in speed and accuracy of our algorithm.

\subsubsection{Profile Likelihood}\label{ssec:likelihood-profile}
We next computed the profile likelihood of the simulated data set. This consists of optimizing the model parameters while keeping one fixed, thus tracing out "profile" which serves as an estimate of the marginal likelihood versus that parameter. We started at the maximum likelihood parameters and stepped through each parameter with 20 points from $-3\sigma$ to $+3\sigma$ based on the parameter uncertainties ($\sigma$) returned at the maximum likelihood computed from the covariance matrix.  We implemented the stepping of the values of each parameter by providing a tight quadratic prior at the grid point, and optimized the likelihood and prior starting at the maximum likelihood or using the previous grid point for initializing each optimization.  The result is the profile of the likelihood maximized over all parameters except the parameter which is fixed at each grid point.

\begin{figure*}
\centering
\includegraphics[width=0.95\textwidth]{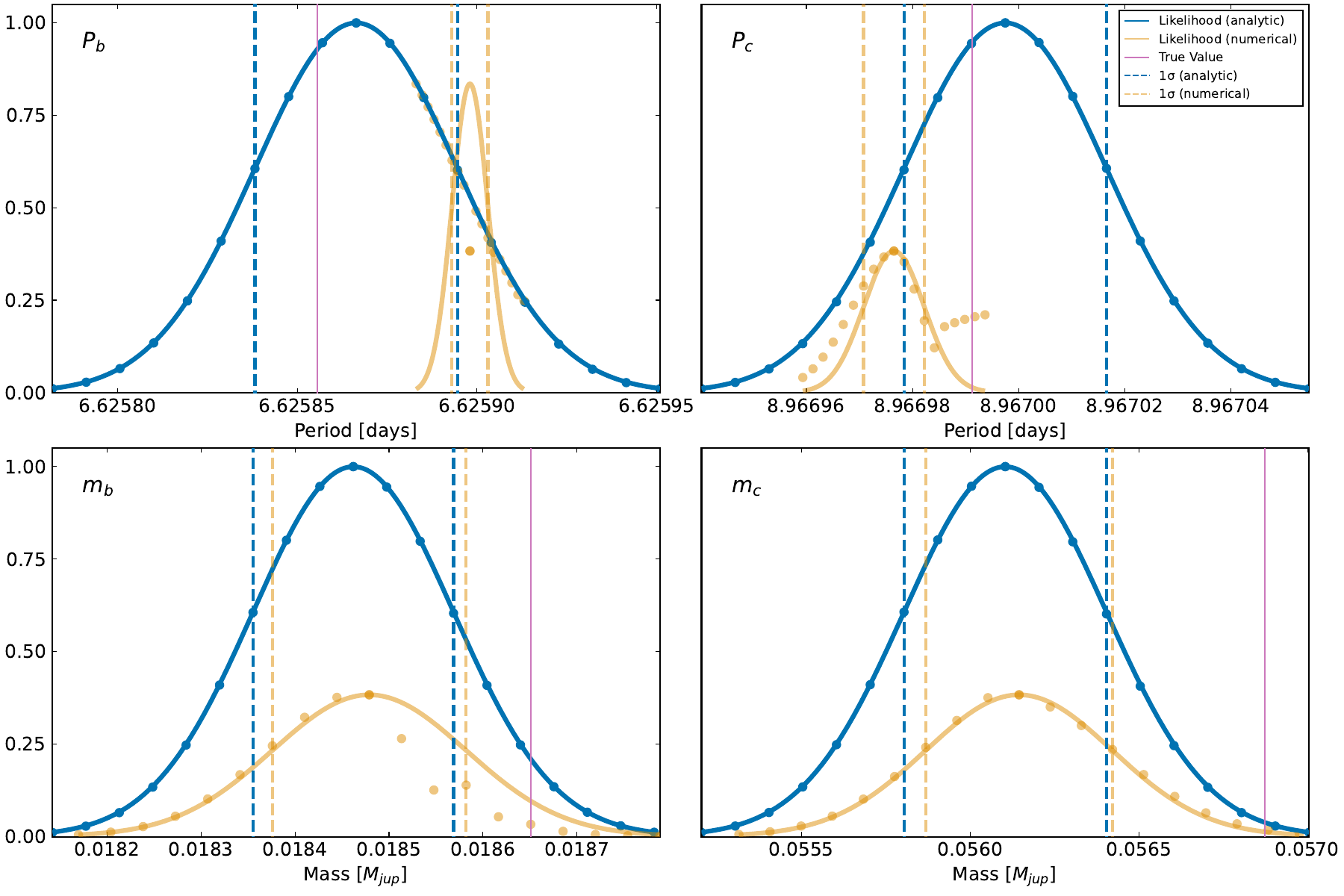}
\caption{Likelihood profile (normalized by the maximum likelihood) versus the labeled model parameters. In each panel, the blue points represent the maximum likelihood value computed while holding the labeled parameter fixed at the given value, the blue lines are a Gaussian function with mean and variance equal to the maximum likelihood value and the variance estimated from the optimization, the orange points and lines are the same as the blue, but computed using numerical derivatives, and the magenta line is the value of the parameter used to generate the synthetic data set. In all cases, the analytic derivative profiles show better agreement with the estimates of the variance from the optimization than the numerical derivatives. }
\label{fig:profile}
\end{figure*}

Figure \ref{fig:profile} shows the profile likelihood for the mass and period of each planet. The blue points represent the results obtained using analytic derivatives, while the orange points use numerical derivatives. Each point is computed using a single optimization run. That is, we do not re-run the optimization if it does not converge on the first pass. As with the initial global optimization, the numerical derivatives perform much more poorly than the analytic derivatives, and the numerical derivatives take much longer to run. Also plotted in the figure are Gaussian profiles using the maximum likelihood and uncertainties derived from the optimization. It is clear that these Gaussian profiles agree extremely well with the profile likelihood, which indicates that the probability distribution is well approximated by a multidimension Gaussian, and it also indicates that the profile likelihood is accurately finding the maximum likelihood subject to the constraint on each parameter. Conversely, the profile likelihood computed with finite-difference derivatives has trouble maximizing the likelihood at each grid point, and doesn't match the Gaussian computed from the optimized likelihood as the optimization did not converge to the global maximum likelihood. Hence, the analytic derivatives provided by our algorithm provide a superior approach to computing the profile likelihood. In practice this means that the maximum likelihood from the numerical derivatives would need to be re-run, and may eventually converge, but with much more computational expense.  In practice, though, a more important test is the efficiency of Bayesian inference of the posterior probability distribution, which we describe next.

\subsection{Posterior inference with Markov Chain Monte Carlo (MCMC)}\label{ssec:mcmc}

Finally, we investigate the potential for faster inference of the posterior probability distribution of the parameters of a photodynamical model. We compare two MCMC sampling methods: 1) Hamiltonian Monte Carlo \citep[HMC][]{Neal2011} and 2) affine-invariant (``AInv") sampling\footnote{\url{https://github.com/madsjulia/AffineInvariantMCMC.jl}.} \citep{Goodman2010} (aka the ``emcee" sampler, which is widely used Python implementation within the field of astronomy, \citealt{ForemanMackey2013}). The particular implementation of HMC that we use is a No U-Turn Sampler \citep[NUTS;][]{Hoffman2014}, which automatically tunes the step size and length of leap-frog integration, yielding a very short correlation length.\footnote{\url{https://github.com/TuringLang/AdvancedHMC.jl}}  

The AInv sampler has several advantages: it is easy to implement, it has no parameters to tune, it does not require derivatives of the posterior with respect to the model parameters, and it can be very efficient for posterior probability distributions which are well-approximated by a multidimensional Gaussian.  Some drawbacks of AInv is that it requires multiple ``walkers" to have their likelihoods evaluated simultaneously at each step, it can perform poorly with posterior probability distributions with non-linear correlations between parameters (including multimodality), and it can perform poorly for higher-dimensional inference, especially degrading beyond $\approx 15$ model parameters due to the diffusive nature of the algorithm, which is a random walk. The computational expense of computing probabilities for simultaneous walkers can be overcome with parallel processing; however, the poor performance for high-dimensional/non-linear posteriors manifests as strongly correlated Markov chains, which causes a higher variance in the posterior probability distribution. In particular, the goal of MCMC is to obtain a large number of independent effective samples to more accurately approximate various integrals or statistics of the posterior probability distribution.  With AInv, in high-dimensional problems, the parameters can be strongly correlated across many Markov chain steps, meaning that statistically independent samples can only be obtained by computing chains for much longer to obtain posterior samples which are many times the correlation length.

The NUTS sampler tries to overcome these drawbacks by adding in a momentum variable for each model parameter with a corresponding kinetic energy term added to the negative log posterior, and then carrying out an integration over this parameter phase-space, at constant probability (i.e. energy), to take much larger steps which are not diffusive, and thus become uncorrelated much more quickly.  The NUTS algorithm also has some advantages and disadvantages.  An advantage is that the correlation length can be greatly reduced relative to standard MCMC;  in fact, it typically provides an independent sample with every Markov chain step, assuming the sampling parameters are tuned.  This advantage comes with several disadvantages: it requires derivatives of the log posterior probability with respect to the model parameters, which can be complicated and more expensive to compute; inaccurate derivatives can cause the energy or probability not to be conserved along a trajectory, which increases the rejection rate; it has numerous parameters which need to be tuned, for which there are a few automated algorithms \citep{Hoffman2014}, but they can be expensive; and each step can require a lengthy symplectic integration in the space of the model parameters and their conjugate momenta, which can be time-consuming.  As these disadvantages can be outweighed by the efficacy of NUTS, one of the main goals of developing a differentiable photodynamical model is to use the NUTS sampler to improve the efficiency of sampling, and so the Photodynamics.jl model addresses this first requirement of HMC/NUTS.\footnote{Note that during the refereeing process for this paper, another paper appeared which computes derivatives of a photodynamical model in JAX using automatic differentiation to compute derivatives \citep[\texttt{jnkepler};][]{Masuda2024}.}

Given the advantages and disadvantages of these two techniques, we make a direct comparison on a set of example photodynamical problems to test which Markov chain inference method provides the greatest sampling efficiency. In particular, we wish to obtain as large a number of statistically-independent ``effective" samples in as little CPU time as possible. 

We compute the correlation length of each parameter and chain, and then divide the product of the total number of chains and steps (after burn-in or adaptation) by the maximum value of the correlation length over all parameters and chains to determine the total number of effective samples.  We then divide the total CPU time taken to run the chains by the number of effective samples to determine the CPU time per effective sample.  We compute this metric for both techniques (AInv and NUTS) to see which method performs the best as a function of the number of free parameters. We also verify that each method produces posterior samples that converge to the same distribution. 

\subsubsection{Synthetic Data}
The previous test problem has a limited number of free parameters for which to make the comparison.  So, instead, we made a comparison for a set of simulated photodynamical models for the TRAPPIST-1 system. We carry out 6 tests on synthetic data, for which we utilize the parameters from \citet{Agol2021}. The first 4 utilize 1600 d of observations with 2 min cadence, for 2, 3, 4, and 5 planets, starting with the inner planets b and c, and adding in the additional planets going outwards. A fifth test uses the outer two planets, g and h, for comparison with the 2-planet b/c test. The final test is for 150 d of observations of all 7 TRAPPIST-1 planets. Each simulated light curve was given noise at the level of JWST NIRSPEC of 51 ppm \citep[e.g.][]{Rathcke2025}. For each planet, we allow the same six parameters to vary as in the test problem, while holding the stellar parameters, planet inclinations, and planet longitudes of ascending node fixed.\footnote{These parameters were held fixed as they can often have an asymmetric or multimodal posterior, which can make sampling less efficient.  We would expect that the affine-invariant sampler will perform worse than the NUTS sampler under these conditions, and so our simulations are conservative.}  We then carry out MCMC inference on these data (12-42 parameters) and compute statistics of the Markov chains for each method. The effective sample sizes (ESS) were calculated with the formulae from \citet{Vehtari2021} using the package \texttt{MCMCChains.jl},\footnote{\url{https://github.com/TuringLang/MCMCChains.jl}} and we took the smaller of the ESS or the total number of samples.  We measure the CPU time per effective sample for each method and each number of planets; the CPU time was tracked with \texttt{CPUTime.jl}.\footnote{\url{https://github.com/schmrlng/CPUTime.jl}}

\subsubsection{MCMC Initialization}
Before starting the Markov chains, we optimized the model parameters using the Levenburg-Marquardt (L-M) algorithm.  We repeated iterations of L-M until the chi-square converged to a minimum value.  We then computed the covariance matrix at the minimum chi-square, which we saved for use in each of the MCMC samplers.  In both cases our priors were uniform, although we placed bounds on the parameters to be physical, such as a positive mass and eccentricity less than one\footnote{In practice, we simply do not include a prior term in our posterior calculations. We assume that our uniform priors cover the region of the parameter space such that the log-posterior is effectively always the log-likelihood function plus a constant.}. For each planet, we sampled the set of parameters: $\{P,t_0,M,e\cos{\omega},e\sin{\omega},k\}$.  We did not use the inverse eccentricity prior \citep{Eastman_2013}, as the eccentricities were well constrained by the data and so this prior should be nearly constant over the range of high posterior probability.

The simulations were carried out for 2500-5000 steps with AInv, with 10\% discarded as burn-in. The number of AInv walkers was three times the number of parameters.  We initialized the walkers by drawing them from a multidimensional Gaussian with mean values given by the maximum likelihood (minimum chi-square) from the initial optimization, and covariance matrix computed at the maximum likelihood.

For the NUTS algorithm, we used the inverse of the covariance matrix as the initial mass matrix.  Note that this is analogous to the geometric transformations recommended by \citet{Tuchow2019}. We carried out 100 adaptation steps for NUTS using STAN's windowed adaptation of the step size and mass matrix \citep{Hoffman2014}.  We set the maximum tree depth to three so that the adaptation does not waste too much time on long trajectories for short steps, and confirm that this choice does not extend the auto-correlation length of the chains. This is then followed by 900 sampling steps. We ran a single NUTS chain for each of the 6 simulated data sets.

\subsubsection{Multiplanet Tests}
\begin{figure}
    \centering
    \includegraphics[width=\linewidth]{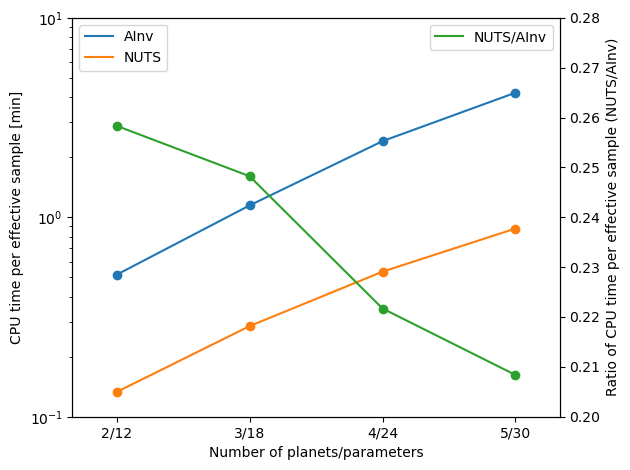}
    \caption{The CPU time per effective sample versus the number of planets/parameters for NUTS (orange) and affine invariant (blue, ``AInv").  The ratio of the CPU times per effective sample versus the number of planets/parameters (green; right axis label).}
    \label{fig:time_per_sample}
\end{figure}

The results of the first 4 simulations are summarized in Figure \ref{fig:time_per_sample} which plots the CPU time per effective sample for the two samplers versus the number of planets/parameters. The CPU time per effective sample for AInv and NUTS grows with the number of parameters/planets.\footnote{These four comparisons are based on simulations carried out in Julia v1.10.3 with an Apple M3 Max processor using a single thread, and with \texttt{BLAS.num\_threads(1)} to limit the linear algebra to a single thread as well.} 

We also plot the ratio of the CPU time per effective sample of NUTS to that of AInv.  For this particular problem, we find that the NUTS algorithm takes about $21-26$\% of the AInv algorithm to acquire an effective sample, or a factor of 4-5 in improvement for NUTS over AInv, where this ratio improves with the number of free parameters. The change with the number of parameters is likely due to the longer correlation lengths of the AInv Markov chains as they execute a random walk in the higher-dimensional parameter space.  We find that the ratio of computation time per effective sample is about the same for planets g and h as it is for planets b and c.  Finally, we reran the AInv chains with multiprocessing using six worker processes and one main process, and found an additional 33\% overhead in this case, which further favors the NUTS sampler by a ratio of 5-7.

\subsubsection{TRAPPIST-1-like Test}
For the 7-planet system, we find a further increase in sampling efficiency for NUTS over AInv. In this test, we aimed to used each of the sampling codes as implemented and attempt to emulate a real-world analysis with 42 free parameters in the model. We run these tests on the computing cluster node described in Section \ref{ssec:comparison}, set the number of available CPU cores to 8 and the total available memory to 16 GB (typical specifications for modern workstation desktop and laptop computers), and set the number of available BLAS threads to the number of available CPU cores. We then run the MCMC sampling as above: 100 adaptation steps with a max tree-depth of 3 and 1000 samples for NUTS, and 2500 samples per 42 $\times$ 3 = 126 walkers for the AInv sampling. 

We find that HMC/NUTS takes 4.12 hours to obtain 1000 effective samples, while AInv sampling takes 4.57 hours to obtain 630 effective samples. So, we would need to run AInv algorithm for another 2.7 hours on 8 cores to achieve the same number of effective samples as HMC/NUTS on a single core. Alternatively, since the HMC/NUTS chains are entirely uncorrelated, we could run 8 HMC/NUTS chains simultaneously and achieve the same number of effective samples in just over 30 min. This yields a ratio of compute time per effective sample of 7\%, or a factor of 14, which is much larger than the ratio we found in the single-threaded cases above (a factor of 4-5) with a smaller number of free parameters.

\section{Summary and Conclusions}\label{sec:conclusions}
Modeling exoplanet transits allows us to place constraints on the properties of the star, the transiting planet, and the planet's orbit. When multiple planets are present, the transit timing variations can provide additional constraints and break degeneracies between the model parameters. Using photodynamics to compute transit light curves allows us to maintain a single, consistent dynamical model -- effectively modeling the photometry of each transit \emph{and} the transit-timing variations, simultaneously.

In this paper, we outlined a novel, analytically differentiable photodynamical model for computing light curves of transiting multiplanet systems. We base the model on previously developed tools: the AHL21 differentiable 4-th order symplectic integrator \citep{AHL2021}, the PK20 impact parameter expansion model \citep{PK2020}, and the differentiable ALFM20 limbdarkening transit model \citep{Agol2020}. This particular composition leads to computationally efficient analytic derivatives of the time-integrated flux with respect to the \textit{N}-body initial conditions and the transit model parameters. Access to the derivatives of the model with respect to the input parameters allows the use of gradient-based Bayesian inference tools such as Hamiltonian Monte Carlo \citep[HMC; ][]{Duane1987,Neal2011,Monnahan2016,Betancourt2017}, which can be more robust for sampling high-dimensional parameter spaces. To the best of our knowledge, this is the first analytically differentiable photodynamical model.

We have implemented the model in the Julia language as \pdjl, which is available on GitHub. Our code compares well with the established \photodynam\footref{fn:photodynam} \citep{Carter2011} code in both accuracy and performance, and is staged to take advantage of the existing \texttt{Julia} ecosystem of data modeling tools. A significant advantage of our code over \photodynam is the computation of accurate derivatives which enable faster optimization, inference, and computation of the information matrix. We have shown that we achieve better posterior sampling efficiency with HMC/NUTS than with the popular Affine-Invariant sampling method for the same model and simulated data. We have made the code open-source to allow others to use and build upon this work.

We emphasize that all of the MCMC results are specific to the particular data sets, models, and computer architectures that we are using here. Many factors can affect the sampling efficiency of MCMC, but this test suite was designed to be a conservative comparison between two well-used sampling schemes: we neglect any parameters that could potentially have multimodal posteriors, and the data sets are high signal-to-noise and so the models are well constrained and the posteriors are well approximated by a multivariate Gaussian -- a regime where AInv is expected to perform very well. However, even in these ideal cases, we find that HMC/NUTS outperforms AInv. It is also worth noting that HMC/NUTS sampling is a serial computation. That is, the sampling does not necessitate a large number of CPU cores to achieve the speed up. Further, if multiple cores \textit{are} available, one could sample multiple chains in parallel with no additional computational overhead.

Since we utilize AHL21 as our dynamical model, this method has the potential for differentiable modeling of arbitrary hierarchical systems. For example: circumbinary planets, exo-moons, or multistar systems. How well the PK20 expansion model performs for these systems has yet to be explored, and computing mutual events and light-travel-time corrections may be more important in these cases. The underlying framework is agnostic to the choice of transit and \textit{N}-body models, provided that they are differentiable either analytically or through automatic differentiation.  Our particular implementation in \nbg currently only includes Newtonian forces between the bodies.  We look forward to users applying this code to other systems or even to problems that we have not anticipated.

\section*{Acknowledgments}

We acknowledge support from the National Science Foundation (NSF) grant AST-1907342, National Aeronautics and Space Administration Nexus for Exoplanet System Science (NASA NExSS) grant No.\ 80NSSC18K0829, and National Aeronautics and Space Administration Exoplanet Research Program (NASA XRP) grant 80NSSC21K1111. Zachary Langford thanks the University of Pennsylvania Fontaine Society for support through the Fontaine Graduate Fellowship. We thank the referee for insightful comments which have improved the paper immensely.  We also thank David Hernandez and Cullen Blake for discussions and comments on the submitted draft of this paper. Zachary Langford acknowledges the use of the University of Pennsylvania General Purpose Cluster, on which some of the computations in this paper were carried out.

\section*{Data Availability}
No new data were generated or analysed in support of this research. Any code used to generate simulated data for the figures will be made available in a GitHub repository. It will be linked to in the main code repository: \url{https://github.com/langfzac/Photodynamics.jl}




\bibliographystyle{mnras}
\bibliography{main} 




\appendix

\section{Dynamical Model Initial Conditions}\label{app:ics}

Here we summarize the initial conditions of the dynamical model, as specified in Cartesian coordinates \citep{Agol2021}, as well as Keplerian orbital elements (\S \ref{app:orbital_elements}). \nbg uses the Cartesian coordinate system to carry out the integration and compute derivatives of the current position with respect to the initial coordinates \citep{AHL2021}. We have also implemented the transform from orbital elements to Cartesian coordinates in \nbg, along with the corresponding derivative transformations. As for \pdjl (\S \ref{sec:implementation}), we include tests against \texttt{BigFloat} finite-difference derivatives to validate the accuracy.

\subsection{Cartesian coordinates} \label{app:cartesian_coordinates}

The Cartesian coordinates utilize a coordinate system for which the sky plane is the $x-y$ plane, while the $z$ axis is along the line of sight, increasing away from the observer. Positions for each body are denoted with a vector $\mathbf{x}_i(t)=\left<x_i(t),y_i(t),z_i(t)\right>^{\intercal}$, while velocities are denoted with $\mathbf{v}_i(t)=\left<\dot x_{i}(t),\dot y_{i}(t),\dot z_{i}(t)\right>^{\intercal}$, with subscript $i=1,..,N$ labelling each body, and $\dot c = \frac{d c}{d t}$ indicates time derivative of variable $c$.  The observer is located at $ \mathbf{x}_{obs} = \left<0,0,-D\right>^{\intercal}$, where $D$ is the distance of the observer to the center of mass of the system (although we don't require the center of mass to be at the origin).  

The initial conditions are completely specified via $\nbq(t_0)$, where $\nbq(t) = \{\mathbf{x}_i(t),\mathbf{v}_i(t), m_i; i=1,...,N\}$.  The vector $\nbq(t)$ has $7N$ elements, where the $7(i-1)+j$th element refers to planet $i$ and the $j$th element of the vector 
\begin{equation}\label{eqn:coordinates_body}
    \nbq_i(t) =\left<x_i(t),y_i(t),z_i(t),\dot x_i(t),\dot y_i(t), \dot z_i(t), m_i\right>^{\intercal}
\end{equation}
where $j=1,...,7$.  Note that we take the origin of the coordinates to be the center of mass of the system, so that a constraint on the initial conditions is $\sum_i m_i q_{i,j}(t_0) = 0$ for $j=1,...,6$, where $q_{i,j}$ denotes the $j$th element of of $\nbq_i(t)$.\footnote{In general, the center-of-mass is allowed to move at a constant velocity, which is not implemented in our initial
conditions, but could be if required.}

\begin{figure*}
\center
	\includegraphics[width= .48\textwidth]{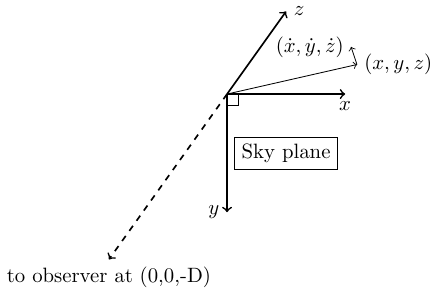}
	\includegraphics[width= .48\textwidth]{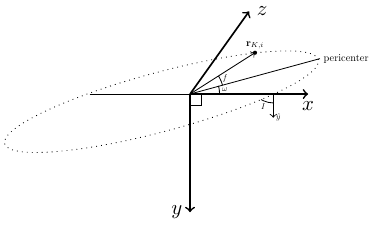}
\caption{Left: Cartesian coordinate system.  Body $i$ is at position $\mathbf{x}_i=\left<x_i, y_i, z_i\right>^{\intercal}$ with velocity $\mathbf{v}_i =\left<\dot x_i, \dot y_i, \dot z_i\right>^{\intercal}$. Right: Orbital elements of the $i$th Keplerian with $\Omega = 0$ (for $\Omega >0$ the orbit rotates about the $z$ axis).  Note that the vector $\mathbf{r}$ extends from the center of mass from the first set of bodies to the center of mass of the second, while in this diagram it is shifted to the origin.
}\label{fig:coordinates}  
\vspace{0.3cm}
\end{figure*}

The coordinate system is right-handed;  with the $x$-axis pointing to the right on the sky, the $y$-axis points downwards, so that $\hat{\mathbf{x}} \times \hat{\mathbf{y}} = \hat{\mathbf{z}}$ points away from the observer, for unit vectors $\{\hat{\mathbf{x}}, \hat{\mathbf{y}}, \hat{\mathbf{z}}\}$ (Figure \ref{fig:coordinates}).

\subsection{Orbital elements} \label{app:orbital_elements}

In most cases we expect that the initial conditions will be specified with instantaneous orbital elements. For this situation we assume that the center-of-mass is stationary, and so we require $N-1$ Keplerians to define the problem. This algorithm is designed with exoplanets in mind, so our plane of reference, the $x-y$ plane, is the sky plane rather than the invariable plane, as in the Solar System.  For transiting exoplanets, the inclinations are close to 90 degrees with respect to the sky plane so that the planets pass in front of the star. However, the $N$-body integrator is applicable to more general $N$-body problems for which differentiation is needed, so the coordinates may be reinterpreted for the problem of choice.

\begin{figure}
    \centering
    \includegraphics{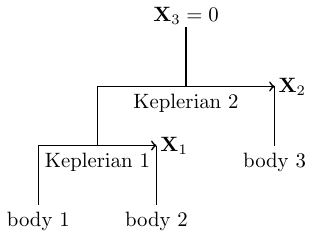}
    \caption{Example of a Keplerian hierarchy initial condition. The first Keplerian consists of the binary orbital motion of the inner two bodies. The second Keplerian consists of the motion of a third body about the center of mass with the first binary. This example corresponds to Jacobi coordinates with three bodies \citep{Wisdom1991}.}
    \label{fig:mobile_diagram}
\end{figure}

We define the initial orbital elements in a hierarchy of Keplerians, where at each level of the hierarchy the instantaneous orbital elements are given at time $t_0$ for the center of mass of one set of bodies orbiting the center of mass of another set of bodies (Figure \ref{fig:mobile_diagram}). We follow the convention of \citet{Hamers2016} in defining the orbital elements and in the conversion of these elements into Cartesian coordinates for the $N$ bodies.

\subsection{Derivatives of initial conditions} \label{app:ics_derivatives}

We expect that most problems will require specifying the orbital elements at an initial time $t_0$.  The $N$-body integrator keeps track of the derivatives with respect to the initial Cartesian coordinates \citep{Agol2021}, and so an additional Jacobian is necessary to transform the derivatives to the initial orbital elements and masses, which we derive in this section.

\subsubsection{Transformation from Keplerian coordinates to Cartesian} \label{app:kep_cart_transform}

\citet{Hamers2016} define an $N\times N$ mass matrix, $\mathbf{A}$, in which the $i$th row corresponds to a single Keplerian and each column to a single body in the system.  In the $i$th row of this matrix, a negative weight is given to the bodies on one side of the $i$th Keplerian, and a positive weight to the bodies on the other side of the Keplerian, with zero weight given to all other bodies.  The final row has a weight for each body which is its fraction of mass times $-1$.  We define ${\bf X}_i$ to be the vector from the center of mass of the first group of bodies to that of the second group for the $i$th Keplerian, and ${\bf X}_N = 0$ is the center-of-mass. With this definition, the transformation from Keplerian to Cartesian coordinates is achieved with
\begin{eqnarray}
    \mathbf{X}_i &=& \sum_{k=1}^{N} A_{i,k} {\bf x}_k,\\ 
    \dot{\bf X}_i &=& \sum_{k=1}^{N} A_{i,k} \dot{\bf x}_k,
\end{eqnarray}
while the inverse transformation is
\begin{eqnarray}
    \mathbf{x}_{n} &=& \sum_{k=1}^{N} A^{-1}_{n,k} {\bf X}_k,\\
    \dot{\bf x}_{n} &=& \sum_{k=1}^{N} A^{-1}_{n,k} \dot{\bf X}_k.
\end{eqnarray}

Each row of the transformation matrix, ${\bf A}$, can be defined with a set of integer indices, $\epsilon_{i,j}$, which labels the $i$th Keplerian, the $j$th body.  For each Keplerian, the first set of bodies have $\epsilon_{ij}=-1$, the second have $\epsilon_{ij}=1$, and the remaining bodies have $\epsilon_{ij}=0$.  Then, the matrix $\mathbf{A}$ is defined as
\begin{eqnarray}\label{eqn:A_matrix}
    A_{ij} = \frac{\epsilon_{ij} m_j}{\sum_k m_k \delta_{\epsilon_{ij}, \epsilon_{ik}}},
\end{eqnarray}
where $\delta_{ij}$ is the Kronecker delta function.

As a concrete example (Figure \ref{fig:mobile_diagram}), for a planetary system with two planets of masses $m_2$ and $m_3$ orbiting a star of mass $m_1$, the $\epsilon$ matrix is
\begin{eqnarray}\label{eqn:epsilon}
    \epsilon = 
        \begin{pmatrix}
            -1 &  1 & 0\\
            -1 & -1 & 1\\
            -1 & -1 & -1
        \end{pmatrix},
\end{eqnarray}
and the matrix ${\bf A}$ is
\begin{eqnarray}
    {\bf A} = 
        \begin{pmatrix}
            -1 &  1 & 0\\
            \frac{-m_1}{m_1+m_2} & \frac{-m_2}{m1+m_2} & 1\\
            \frac{-m_1}{m_1+m_2+m_3} & \frac{-m_2}{m_1+m_2+m_3} & \frac{-m_3}{m_1+m_2+m_3}
        \end{pmatrix}.
\end{eqnarray}

The sum of the masses participating in the $i$th Keplerian is given by 
\begin{eqnarray}\label{eqn:mass_keplerian}
    M_i = \sum_j \vert \epsilon_{i,j}\vert m_j.
\end{eqnarray}

Note that for $\epsilon$ and $\mathbf{A}$ we have chosen the opposite sign convention as \citet{Hamers2016};  e.g., for the inner pair in this example we define the Keplerian coordinate as $\mathbf{X}_1 = \mathbf{x}_2 - \mathbf{x}_1$, while \citet{Hamers2016} define it as $\mathbf{X}_1 =  \mathbf{x}_1 - \mathbf{x}_2$.  We prefer our definition as it indicates that the inner planet orbits the star (although in fact they both orbit their center of mass).

\subsubsection{Derivative of initial Cartesian coordinates with respect to Keplerian orbital elements} \label{app:Keplerian_derivatives}

We define each Keplerian in the hierarchy by the mass sum, $M_i$, along with a set of orbital elements,
\begin{equation}\label{eqn:elements_hierarchy}
    \boldsymbol{\eta}_i = \left\{P_i,\tau_i,k_i,h_i,I_i,\Omega_i\right\},
\end{equation}

where $P_i$ is the orbital period, $\tau_i$ is the time of inferior conjunction (or time of transit in the edge-on planetary case), $k_i = e_i \cos{\omega_i}$, $h_i=e_i\sin{\omega_i}$,

for eccentricity $e_i$ and argument of periastron $\omega_i$, $I_i$ the orbital inclination, and $\Omega_i$ the longitude of ascending node.  This is the basis set of variables for our initial conditions.

We follow the notation of \cite{Murray1999} (chapter 2) in transforming from orbital elements, to the Cartesian coordinates, ${\bf X}_i$ of each Keplerian:
\begin{eqnarray}\label{eqn:initial_Keplerian}
    {\bf X}_i &=& \begin{pmatrix}
        X_i \\ Y_i \\ Z_i
    \end{pmatrix} = {\bf P}_{i}  {\mathbfcal X}_i,\\
    {\mathbfcal X}_i &=& a_i \begin{pmatrix}
        \cos{E_i} - e_i \\ \sqrt{1-e_i^2} \sin{E_i} \\ 0
    \end{pmatrix},
\end{eqnarray}
where the vector ${\mathbfcal X}_i$ contains the orbital coordinates in the orbital plane aligned with pericenter oriented along the semi-major axis, $a_i$ is the semi-major axis, $E_i$ is the eccentric anomaly, and $e_i$ is the eccentricity.  The rotation matrix ${\bf P}_i$ is defined below.   
 
To solve for the eccentric anomaly requires Kepler's equation, given by
\begin{eqnarray}\label{eqn:ecc_meanmotion}
    {\cal M}_i = E_i + e_i \sin{E_i},
\end{eqnarray}
where ${\cal M}_i$ is the mean anomaly. We solve Kepler's equation with a standard solver \citep{Murray1999}, which uses the method of \citet{Danby1983}, consisting of a higher order Newton method with an initial guess of $E_i = \mathcal{M}_i + 0.85 e_i \text{sgn}(\sin(\mathcal{M}_i))$.

Note that although the derivatives of the initial Keplerian Cartesian coordinates are computed with respect to the orbital elements of each Keplerian in the hierarchy, $\boldsymbol{\eta}_i$, and the masses, $M_i$, we find it convenient to take derivatives with respect to several intermediate functions, $e_i$, $a_i$, $E_i$, ${\cal M}_i$, $t_{\mathrm{p},i}$ and $\theta_i$, which are in turn a function of the orbital elements and mass of the $i$th Keplerian.   We then apply the chain rule to these to obtain the derivatives with respect to the orbital elements and masses. Kepler's equation defines $E_i$ implicitly in terms of ${\cal M}_i$ and $e_i$, from which we obtain the partial derivatives of $E_i({\cal M}_i,e_i)$ with respect to ${\cal M}_i$ and $e_i$, given below. We define the time of periastron passage, $t_{\mathrm{p},i}$, and an argument used in computing it, $\theta_i$, as intermediate functions. These and the other intermediate functions are defined as:
\begin{eqnarray}\label{eqn:a_tp}
    {\cal M}_i(P_i,t_{\mathrm{p},i}) &=& \frac{2\pi}{P_i} (t_0 - t_{\mathrm{p},i})\\
    e_i(k_i,h_i) &=& \sqrt{k_i^2+h_i^2}\\
    a_i(M_i,P_i) &=& \left[\frac{G M_i P_i^2}{4\pi^2}\right]^{1/3},
\end{eqnarray}
\begin{eqnarray}
    &&t_{\mathrm{p},i}(P_i,\tau_i,e_i(k_i,h_i),\theta_i(e_i(k_i,h_i),k_i,h_i),k_i,h_i) \cr 
    &=& \tau_i -\sqrt{1-e_i^2} \frac{P_i}{2\pi} \\
    &\times& \left[
    \frac{k_i}{1-h_i} +
    2 (1-e_i^2)^{-1/2} \tan^{-1} \theta_i\right],
\end{eqnarray}
\begin{eqnarray} \label{eqn:thetai}
    \theta_i(e_i(k_i,h_i),k_i,h_i) = \left(\frac{1-e_i}{1+e_i}\right)^{1/2} \frac{h_i+k_i + e_i}{h_i-k_i-e_i},
\end{eqnarray}
where $G$ is Newton's constant.  Note that although the mean anomaly is defined in terms of $t_0$, we don't include it as an argument as it is held fixed, and so we don't require partial derivatives. 

We use these intermediate functions throughout the computation for more compact expressions and efficient code.  We use partial derivatives, e.g.\ $\frac{\partial {\cal M}}{\partial t_{\mathrm{p},i}}$, to denote differentiation with respect to these intermediate quantities, and full derivatives for differentiation of these intermediate functions with respect to the orbital elements and masses, e.g.\ $\frac{d t_{\mathrm{p},i}}{d k_i}$, and we apply the chain rule to obtain the full derivatives for the Cartesian coordinates with respect to the orbital elements and masses below (equations \ref{eqn:initial_condition_full_derivative_start} - \ref{eqn:initial_condition_full_derivative_end}).

To proceed, we need the rotation matrix for the $i$th Keplerian, ${\bf P}_i$, which is given by
\begin{eqnarray}\label{eqn:rotation_matrix}
    {\bf P}_i &=& {\bf P}_{3,i} {\bf P}_{2,i} {\bf P}_{1,i},
\end{eqnarray}
which is the product of three rotation matrices
\begin{eqnarray}\label{eqn:rotation_matrix123}
    {\bf P}_{1,i}(e_i(k_i,h_i),k_i,h_i) &=& \frac{1}{e_i}\begin{pmatrix}
        k_i & -h_i & 0\\
        h_i & k_i &  0\\
        0 & 0 & 1
    \end{pmatrix},\\
    {\bf P}_{2,i}(I_i) &=& \begin{pmatrix}
        1 & 0 & 0 \\
        0 & \cos{I_i} & -\sin{I_i}\\
        0 & \sin{I_i} &  \cos{I_i}
    \end{pmatrix},\\
    {\bf P}_{3,i}(\Omega_i) &=& \begin{pmatrix}
        \cos{\Omega_i} & -\sin{\Omega_i} & 0 \\
        \sin{\Omega_i} & \cos{\Omega_i} & 0\\
        0 & 0 &  1
    \end{pmatrix}.
\end{eqnarray}
and below we use the matrix products:
\begin{eqnarray}\label{eqn:rotation_products}
    {\bf P}_{32,i}  &=& {\bf P}_{3,i} {\bf P}_{2,i},\\
    {\bf P}_{21,i}  &=& {\bf P}_{2,i} {\bf P}_{1,i}.
\end{eqnarray}

With these definitions, the corresponding derivatives are given by 
\begin{eqnarray}
    \frac{\partial {\bf X}_i}{\partial a_i} &=& \frac{1}{a_i} {\bf X}_i,\\
    \frac{\partial {\bf X}_i}{\partial E_i} &=& {\bf P}_i a_i \begin{pmatrix}
        -\sin{E_i}\\ \sqrt{1-e_i^2}\cos{E_i} \\ 0
    \end{pmatrix},
\end{eqnarray}
\begin{eqnarray}
    \frac{\partial {\bf X}_i}{\partial k_i} &=& {\bf P}_{32,i}  \frac{1}{e_i} {\mathbfcal X}_i,\\
    \frac{\partial {\bf X}_i}{\partial h_i} &=& {\bf P}_{32,i}  \frac{a_i}{e_i} \begin{pmatrix}
        -\sqrt{1-e_i^2} \sin{E_i} \\ \cos{E_i} - e_i \\ 0 
    \end{pmatrix},
\end{eqnarray}
\begin{eqnarray}
    \frac{\partial {\bf X}_i}{\partial e_i} &=& -\frac{1}{e_i}{\bf X}_i + {\bf P}_i a_i \begin{pmatrix}
        -1 \\ -\frac{e_i}{\sqrt{1-e_i^2}} \sin{E_i}\\ 0
    \end{pmatrix}.
\end{eqnarray}

The initial velocities of the Keplerians are given by
\begin{eqnarray}
    \dot {\bf X}_i &=& \begin{pmatrix}
        \dot X_i \\ \dot Y_i \\ \dot Z_i
    \end{pmatrix}= {\bf P}_i  \dot {\mathbfcal X}_i,
\end{eqnarray}
with
\begin{eqnarray}\label{eqn:keplerian_Xnr}
    \dot {\mathbfcal X}_i &=& \frac{n_i a_i^2}{r_i} \begin{pmatrix}
        -\sin{E_i} \\ \sqrt{1-e_i^2} \cos{E_i} \\ 0
    \end{pmatrix},
\end{eqnarray}
where
\begin{eqnarray}
    n_i &=& \frac{2\pi}{P_i},\\
    r_i &=& a_i(1-e_i \cos{E_i}),
\end{eqnarray}

with derivatives 
\begin{eqnarray}
    \frac{\partial \dot {\bf X}_i}{\partial a_i} &=& \frac{1}{a_i} \dot {\bf X}_i,\\
    \frac{\partial \dot {\bf X}_i}{\partial P_i} &=& -\frac{1}{P_i} \dot {\bf X}_i,
\end{eqnarray}
\begin{eqnarray}
    \frac{\partial \dot{\bf X}_i}{\partial E_i} &=& -\dot {\bf X}_i \frac{e_i \sin{E_i}}{1-e_i \cos{E_i}} \notag \\ 
    &-& {\bf P}_{i} \frac{n_i a_i^2}{r_i}
    \begin{pmatrix}
        \cos{E_i} \\ \sqrt{1-e_i^2} \sin{E_i} \\0
    \end{pmatrix},
\end{eqnarray}

\begin{eqnarray}
    \frac{\partial \dot{\bf X}_i}{\partial e_i} &=& -\frac{1}{e_i} \dot {\bf X}_i  + \dot{\bf X}_i \frac{\cos{E_i}}{1-e_i\cos{E_i}}\notag \\
    &+& {\bf P}_i \frac{n_i a_i^2}{r_i} \begin{pmatrix}
        0 \\ \frac{e_i}{\sqrt{1-e_i^2}}\cos{E_i}\\ 0
    \end{pmatrix},
\end{eqnarray}
and finally, 

\begin{eqnarray}
    \frac{\partial \dot{\bf X}_i}{\partial k_i} &=& {\bf P}_{32,i}  \frac{1}{e_i} \dot {\mathbfcal X}_i,\\
    \frac{\partial \dot{\bf X}_i}{\partial h_i} &=& {\bf P}_{32,i}  \frac{n_i a_i^2}{e_ir_i} 
    \begin{pmatrix}
        -\sqrt{1-e_i^2} \cos{E_i} \\ -\sin{E_i} \\0
    \end{pmatrix}.
\end{eqnarray}

Some additional partial derivatives are required to complete the differentiation with respect to the orbital elements and masses since $a_i$ is a function $a_i(P_i,M_i)$:
\begin{eqnarray}
    \frac{\partial a_i}{\partial P_i} &=& \frac{2a_i}{3P_i},\\
    \frac{\partial a_i}{\partial M_i} &=& \frac{a_i}{3M_i};
\end{eqnarray}
$e_i$ is a function $e_i(k_i,h_i)$:
\begin{eqnarray}
    \frac{\partial e_i}{\partial k_i} &=& \frac{k_i}{e_i},\\
    \frac{\partial e_i}{\partial h_i} &=& \frac{h_i}{e_i},
\end{eqnarray}
and the intermediate variable, $\theta_i$ (equation \ref{eqn:thetai})
is a function  $\theta_i(e_i(k_i,h_i),k_i,h_i)$: 
\begin{eqnarray}
    \frac{\partial \theta_i}{\partial e_i} &=& \frac{(e_i+k_i)^2+2(1-e_i^2)h_i-h_i^2}{\sqrt{1-e_i^2}(1+e_i)(h_i-k_i-e_i)^2},\\
    \frac{d \theta_i}{d k_i} &=& \frac{\partial \theta_i}{\partial e_i} \frac{k_i}{e_i}+2\sqrt{\frac{1-e_i}{1+e_i}} \frac{h_i}{(h_i-k_i-e_i)^2},\\
    \frac{d \theta_i}{d h_i} &=& \frac{\partial \theta_i}{\partial e_i} \frac{h_i}{e_i}-2\sqrt{\frac{1-e_i}{1+e_i}} \frac{k_i+e_i}{(h_i-k_i-e_i)^2};
\end{eqnarray}
$t_{\mathrm{p},i}$ is a function  $t_{\mathrm{p},i}(P_i,\tau_i,e_i(k_i,h_i),\theta_i(e_i(k_i,h_i),k_i,h_i),k_i,h_i)$,
\begin{eqnarray}
    \frac{\partial  t_{\mathrm{p},i}}{\partial P_i} &=& \frac{t_{\mathrm{p},i}-\tau_i}{P_i}\\
    \frac{\partial  t_{\mathrm{p},i}}{\partial \tau_i} &=& 1,\\
    \frac{\partial t_{\mathrm{p},i}}{\partial e_i} &=& -\frac{e_i}{1-e_i^2}(t_{\mathrm{p},i}-t_0) - \psi_i \frac{\partial \theta_i}{\partial e_i},\\
    \frac{d t_{\mathrm{p},i}}{d k_i} &=& \frac{\partial t_{\mathrm{p},i}}{\partial e_i}\frac{k_i}{e_i} -\sqrt{1-e_i^2}\frac{P_i}{2\pi(1-h_i)}-\psi_i\frac{d \theta_i}{d k_i},\\
    \frac{d t_{\mathrm{p},i}}{d h_i} &=& \frac{\partial t_{\mathrm{p},i}}{\partial e_i}\frac{h_i}{e_i}-\sqrt{1-e_i^2}\frac{P_ik_i}{2\pi(1-h_i)^2}-\psi_i \frac{d \theta_i}{d h_i},
\end{eqnarray}
where for compactness of notation we define
\begin{eqnarray}\label{eqn:psi}
    \psi_i &=& \frac{P_i}{\pi(1+\theta_i^2)};
\end{eqnarray}
$\mathcal{M}_i$ is a function  $\mathcal{M}_i(P_i, t_{\mathrm{p},i})$ (note that $t_0$ is the starting time of integration is fixed),
\begin{eqnarray}
    \frac{\partial {\cal M}_i}{\partial P_i} &=& -\frac{{\cal M}_i}{P_i},\\
    \frac{\partial {\cal M}_i}{\partial t_{\mathrm{p},i}} &=& -n_i,\\
    \frac{d {\cal M}_i}{d P_i} &=& \frac{\partial {\cal M}_i}{\partial P_i} - n_i \frac{\partial  t_{\mathrm{p},i}}{\partial P_i};
\end{eqnarray}
$E_i$ is a function  $E_i({\cal M}_i,e_i(k_i,h_i))$,
\begin{eqnarray}
    \frac{\partial E_i}{\partial e_i} &=& \frac{\sin{E_i}}{1-e_i \cos{E_i}},\\
    \frac{\partial E_i}{\partial {\cal M}_i} &=& \frac{1}{1-e_i\cos{E_i}},
\end{eqnarray}
so that
\begin{eqnarray}
    \frac{d E_i}{d k_i} &=& \frac{\sin{E_i}}{1-e_i \cos{E_i}}\frac{k_i}{e_i} +\frac{\partial E_i}{\partial {\cal M}_i}\frac{\partial {\cal M}_i}{\partial t_{\mathrm{p},i}} \frac{d t_{\mathrm{p},i}}{d k_i},\\
    \frac{d E_i}{d h_i} &=& \frac{\sin{E_i}}{1-e_i \cos{E_i}}\frac{h_i}{e_i} + \frac{\partial E_i}{\partial {\cal M}_i}\frac{\partial {\cal M}_i}{\partial t_{\mathrm{p},i}} \frac{d t_{\mathrm{p},i}}{d h_i}.
\end{eqnarray}

With these definitions, the final expressions for the derivatives are
\begin{eqnarray} 
    \frac{d \mathbf{X}_i}{d P_i} &=& \frac{\partial \mathbf{X}_i}{\partial a_i}
    \frac{\partial a_i}{\partial P_i}+ \frac{\partial \mathbf{X}_i}{\partial E_i} \frac{\partial E_i}{\partial {\cal M}_i} \frac{d {\cal M}_i}{d P_i}, \label{eqn:initial_condition_full_derivative_start}\\
    \frac{d \mathbf{X}_i}{d \tau_i} &=& \frac{\partial \mathbf{X}_i}{\partial E_i} \frac{\partial E_i}{\partial {\cal M}_i}\frac{\partial {\cal M}_i}{\partial t_{\mathrm{p},i}} \frac{\partial t_{\mathrm{p},i}}{\partial \tau_i},\\ 
    \frac{d \mathbf{X}_i}{d k_i} &=& \frac{\partial \mathbf{X}_i}{\partial k_i} + \frac{\partial {\bf X}_i}{\partial e_i}\frac{k_i}{e_i} +\frac{\partial \mathbf{X}_i}{\partial E_i} \frac{d E_i}{d k_i},\\
    \frac{d \mathbf{X}_i}{d h_i} &=& \frac{\partial \mathbf{X}_i}{\partial h_i} + \frac{\partial {\bf X}_i}{\partial e_i}\frac{h_i}{e_i} +\frac{\partial \mathbf{X}_i}{\partial E_i}  \frac{d E_i}{d h_i},\\ 
    \frac{d {\bf X}_i}{d I_i} &=& {\bf P}_{3,i}
    \begin{pmatrix}
        0 & 0 & 0\\ 
        0 & -\sin{I_i} & -\cos{I_i} \\ 
        0 & \cos{I_i} & -\sin{I_i}
    \end{pmatrix}
    {\bf P}_{1,i} {\mathbfcal X}_i,\\
    \frac{d {\bf X}_i}{d \Omega_i} &=& 
    \begin{pmatrix}
        -\sin{\Omega_i} & -\cos{\Omega_i} & 0\\
        \cos{\Omega_i} & -\sin{\Omega_i} & 0 \\ 
        0 & 0 & 0
    \end{pmatrix} {\bf P}_{21,i}  {\mathbfcal X}_i,\\
    \frac{d \mathbf{X}_i}{d M_i} &=& \frac{\partial \mathbf{X}_i}{\partial a_i}\frac{\partial a_i}{\partial M_i},
\end{eqnarray}
and
\begin{eqnarray}
    \frac{d \dot{\bf X}_i}{d P_i} & = & \frac{\partial \dot{\bf X}_i}{\partial P_i}+\frac{\partial \dot{\bf X}_i}{\partial a_i}\frac{\partial a_i}{\partial P_i}+ \frac{\partial \dot{\bf X}_i}{\partial E_i} \frac{\partial E_i}{\partial {\cal M}_i} \frac{d {\cal M}_i}{d P_i},\\
    \frac{d \dot{\bf X}_i}{d \tau_i} &=& \frac{\partial \dot{\bf X}_i}{\partial E_i} \frac{\partial E_i}{\partial {\cal M}_i}\frac{\partial {\cal M}_i}{\partial t_{\mathrm{p},i}} \frac{\partial t_{\mathrm{p},i}}{\partial \tau_i},\\ 
    \frac{d \dot{\bf X}_i}{d k_i} &=& \frac{\partial \dot{\bf X}_i}{\partial k_i} + \frac{\partial \dot{\bf X}_i}{\partial e_i} \frac{k_i}{e_i} + \frac{\partial \dot{\bf X}_i}{\partial E_i} \frac{d E_i}{d k_i},\\
    \frac{d \dot{\bf X}_i}{d h_i} &=& \frac{\partial \dot{\bf X}_i}{\partial h_i} + \frac{\partial \dot{\bf X}_i}{\partial e_i} \frac{h_i}{e_i} + \frac{\partial \dot{\bf X}_i}{\partial E_i} \frac{d E_i}{d h_i},\\ 
    \frac{d \dot{\bf X}_i}{d I_i} &=& {\bf P}_{3,i} \begin{pmatrix}
        0 & 0 & 0 \\
        0 & -\sin{I_i} & -\cos{I_i}\\ 
        0 & \cos{I_i} & -\sin{I_i}
    \end{pmatrix} {\bf P}_1  \dot {\mathbfcal X}_i,\\
    \frac{d \dot {\bf X}_i}{d \Omega_i} &=& 
    \begin{pmatrix}
        -\sin{\Omega_i} & -\cos{\Omega_i} & 0\\ \cos{\Omega_i} & -\sin{\Omega_i} & 0\\ 
        0 & 0 & 0
    \end{pmatrix} {\bf P}_{21,i}  \dot {\mathbfcal X}_i,\\
    \frac{d \dot{\bf X}_i}{d M_i} &=& \frac{\partial \dot{\bf X}_i}{\partial a_i}\frac{\partial a_i}{\partial M_i}.  \label{eqn:initial_condition_full_derivative_end}
\end{eqnarray}

We subsequently refer to these derivatives as $\frac{d {\bf X}_i}{d \boldsymbol{\eta}_i}$, $\frac{d \dot{\bf X}_i}{d \boldsymbol{\eta}_i}$, $\frac{d {\bf X}_i}{d M_i}$, and $\frac{d \dot{\bf X}_i}{d M_i}$. This completes the definition of the coordinates of each hierarchical Keplerian in terms of the initial orbital elements;  now we turn to converting the initial hierarchical coordinates to the initial Cartesian coordinates of each body at time $t_0$.

\subsubsection{Propagating derivatives from initial Keplerian coordinates to center-of-mass Cartesian coordinates}

With the initial Keplerian coordinates, and derivatives with respect to orbital elements, defined in \S \ref{app:Keplerian_derivatives}, and the transformation from Keplerian coordinates to Cartesian coordinates in \S \ref{app:kep_cart_transform}, we can now use the chain rule to compute the derivatives of the initial Cartesian coordinates with
respect to the orbital elements.

The transformation matrix, ${\bf A}$, only depends on the masses of the bodies, $m_k$, and so the derivatives are given by 
\begin{eqnarray}
    \frac{\partial A_{ij}}{\partial m_k}  = \frac{\delta_{kj}\epsilon_{ij}}{\sum_l m_l \delta_{\epsilon_{ij}, \epsilon_{il}}} -  \frac{\delta_{\epsilon_{ij},\epsilon_{ik}}\epsilon_{ij}m_j}{\left(\sum_l m_l \delta_{\epsilon_{ij}, \epsilon_{il}}\right)^2}.
\end{eqnarray}

The derivative of the inverse of ${\bf A}$ is given by
\begin{eqnarray}
    \frac{\partial {\bf A}^{-1}}{\partial m_k}  = - {\bf A}^{-1} \frac{\partial {\bf A}}{\partial m_k} {\bf A}^{-1},
\end{eqnarray}
which follows from $({\bf A} {\bf A}^{-1})^\prime = I^\prime = 0$.

The derivatives of the initial Cartesian coordinates with respect to the Keplerian orbital elements is given by:
\begin{eqnarray}
    \frac{\partial {\bf x}_n}{\partial \boldsymbol{\eta}_i} &=& \sum_{k=1}^{N} A^{-1}_{n,k} \frac{\partial {\bf X}_k}{\partial \boldsymbol{\eta}_i},\\
    \frac{\partial \dot{\bf x}_n}{\partial \boldsymbol{\eta}_i} &=& \sum_{k=1}^{N} A^{-1}_{n,k} \frac{\partial \dot{\bf X}_k}{\partial \boldsymbol{\eta}_i}.
\end{eqnarray}

We treat the masses separately as ${\bf A}$ depends upon the body masses, giving:
\begin{eqnarray}
    \frac{\partial {\bf x}_n}{\partial m_i} &=& \sum_{k=1}^{N} \left[ A^{-1}_{n,k} \vert \epsilon_{k,i}\vert \frac{\partial {\bf X}_k}{\partial M_k} + \frac{\partial A^{-1}_{n,k}}{\partial m_i} {\bf X}_k\right],  \\
    \frac{\partial \dot{\bf x}_n}{\partial m_i} &=& \sum_{k=1}^{N} \left[ A^{-1}_{n,k} \vert \epsilon_{k,i}\vert \frac{\partial \dot{\bf X}_k}{\partial M_k} +\frac{\partial A^{-1}_{n,k}}{\partial m_i} \dot{\bf X}_k\right].
\end{eqnarray}

We place these derivatives into a Jacobian matrix, ${\bf J}_{\mathrm{init}}$, which is $7N \times 7N$ in size, given by
\begin{eqnarray}\label{eqn:Jinit}
    {\bf J}_{\mathrm{init}} = \frac{\partial {\bf q(t_0)}}{\partial \boldsymbol{\eta}},
\end{eqnarray}
where $\boldsymbol{\eta}$
contains the $N$ masses and $6(N-1)$ initial orbital elements.


\bsp	
\label{lastpage}
\end{document}